\documentclass[10p]{article}
\usepackage[T1]{fontenc}
\usepackage{lmodern}

\usepackage[english]{babel}

\usepackage[letterpaper,top=3cm,bottom=3.5cm,left=3cm,right=3cm,marginparwidth=1.75cm]{geometry}

\usepackage{float}

\usepackage{color}
\usepackage[font=footnotesize, skip=8pt]{caption}
\usepackage{subcaption}

\usepackage{multirow}
\usepackage{tikz}
\usetikzlibrary{shapes.geometric, arrows, fit, decorations.pathreplacing, arrows.meta}

\usepackage{nicematrix}
\usepackage{makecell} 

\usepackage{titlesec} 

\usepackage{array}
\newcolumntype{L}[1]{>{\raggedright\let\newline\\\arraybackslash\hspace{0pt}}m{#1}}
\newcolumntype{C}[1]{>{\centering\let\newline\\\arraybackslash\hspace{0pt}}m{#1}}
\newcolumntype{R}[1]{>{\raggedleft\let\newline\\\arraybackslash\hspace{0pt}}m{#1}}

\usepackage[style = nature, autocite=superscript, backend=bibtex, giveninits=true, isbn=false, doi=false, eprint=false]{biblatex}
\DeclareFieldFormat{pages}{#1} 
\renewbibmacro{in:}{%
	\ifentrytype{article}
	{}
	{\bibstring{in}%
		\printunit{\intitlepunct}}}
\usepackage{amssymb}
\usepackage{amsmath}
\usepackage{graphicx}
\usepackage{hyperref}
\usepackage[capitalise]{cleveref}

\newcommand{\unnumsubsection}[2]{%
		\phantomsection
	\subsection*{#1}%
	\addcontentsline{toc}{subsection}{#1}%
\label{#2}%
}

\newcommand{\unnumsubsubsection}[2]{%
	\phantomsection
	\subsubsection*{#1}%
	\label{#2}%
}

\titleformat{\subsubsection}
{\normalfont\fontsize{11pt}{13pt}\selectfont\bfseries}
{\thesubsubsection}{1em}{}

\makeatletter
\renewcommand*{\@fnsymbol}[1]{\@arabic{#1}}
\renewenvironment{abstract}{
	\begin{center}
		\begin{minipage}{0.85\linewidth}  
			\setlength{\parindent}{0pt}       
			\ignorespaces
		}{
		\end{minipage}
	\end{center}
	\vspace{1.5em}
}
\makeatletter
\patchcmd{\@maketitle}{\begin{center}}{\vspace{-20pt}\begin{center}}{}{}
\makeatother

\title{\Huge \textbf{An introduction to memory competitions, records and techniques}}
\author{Bastian Wiederhold\\
	\small\texttt{bastian.wiederhold@lmu.de}}
\date{
	\small Bernstein Center for Computational Neuroscience\\
	and Faculty of Biology, Ludwig-Maximilians-Universit\"at, Munich\\
	\small\today}

\begin{document}


\maketitle

\begin{abstract} 
Anyone who has tried to memorize a one-hundred-digit number can attest that the human brain acquires abstract information slowly. However, following a three-decade increase of results at memory competitions, the best competitors manage to memorize a one-hundred-digit number in under $15$ seconds. This documents explores the origins of this trend: how competitions, records and techniques have evolved. In the process several phenomena are discussed: the actual process of memorization is likely even faster, records are governed by a power law, numbers are easier than playing cards and associations are at the core of memory.
\end{abstract}


Memory techniques have been known since antiquity \autocite{ancientmnemonics}, but only over the past three decades memory competitions have tested the limits of mnemonic strategies. The most famous is the ``memory palace'' or ``method of loci''. During memorization, competitors first transform information into ``mnemonic images'': memorable objects, people or scenes. These images are then linked to prominent locations along a mental walk. During recall, competitors reimagine the same mental journey and decode the mnemonic images to retrieve the stored information. Optimization of these strategies lead to a dramatic increase of performances. For example, the record for memorizing the sequence of cards in shuffled deck of $52$ playing cards was two minutes in $1993$ and is now $12.74$ seconds by Shijir-Erdene Bat-Enkh set in $2018$ \autocite{IAMrecords}. Intuitively, memory palaces and other techniques work because memorizing events, objects, people, emotions and locations coincides much more with human reality than pure abstract information, leading to higher retention. As such, we could interpret mnemonic techniques as a way to demonstrate the speed of human memory in its intended environment under training.

We analyze scores from four competition formats. First, the famous $\pi$ competitions, in which the objective is to recite as many digits of the irrational number $\pi$ as possible \autocite{pi}. Second, the classical decathlon with disciplines based on digits, playing cards, words, images and names and  memorization times spanning several minutes up to hours \autocite{Rulebook}. Third, the modern format Memory League (ML), which has condensed the ten disciplines of the decathlon into just six: digits, playing cards, words, images, national and international names \autocite{MemoryLeague}. The memorization time lasts a maximum of one minute and focus is on one-vs.-one competing. Fourth, speed-memory.com, which features the shortest disciplines of one and four seconds decimal- and binary-digit memorization \autocite{speed_memory}. Altogether, the competitions cover the whole range of possible memorization times.

To compare different disciplines, we calculated the minimal information processing rate in bit/s necessary to explain top performances. Essentially, we convert the ``content'' of a discipline into the number of $0$ or $1$ options, a computer would need to save every second, to perform in the same way as the record-setting competitor. The information rates allow to compare seemingly unrelated disciplines such as memorizing digits and the names of people. Fascinatingly, given a similar time to memorize, competitors manage to store a largely comparable amount of information, regardless of the disciplines.
%
The scientific aspects of this work are condensed in a companion work \autocite{memory_article}.

\setcounter{tocdepth}{2}
\tableofcontents

\section{Short guide to memory competitions}

\unnumsubsection{\texorpdfstring{$\pi$}{pi} competitions}{}
The most famous irrational number $\pi$ has a decimal expansion without recurring sequences, which has long fascinated humanity. Memorizing $\pi$ is thus a classic challenge even predating the advent of organized memory competitions \autocite{pi}. Without time constraints, no errors allowed and hours of memorization and recall, $\pi$-competitions are also a test of nerves, concentration and endurance. 

\unnumsubsection{Classical format}{subsec:classical_format}
Modern memory competitions started with the first world championship in $1991$, which initiated the classical format: a mental decathlon featuring the memorization of decimal, binary and auditive digits, cards, words, images, names and historic dates (\Cref{table:classical_disciplines}).
Competitions take place over the course of one, two or three days with correspondingly longer disciplines; only the world championship lasts three days. Disciplines feature a memorization period lasting from $5$ to $60$ minutes, followed by a recall phase two or three times as long. The overall winner is determined by a points system similar to the the track-and-field decathlon \autocite{Rulebook}. 
In $2016$, the International Association of Memory (IAM) was founded, splitting from the World Memory Sports Council (WMSC), which hosted all previous classical competitions. This analysis will use the data provided by the IAM \autocite{IAMstatistics}. 

\unnumsubsection{Memory League}{}
The launch of Memory League (ML) in $2015$, first under the name Extreme Memory Tournament \autocite{MemoryLeague}, accelerated the transition to digital competitions. ML features fewer disciplines with a maximum of one minute memorization and four minutes recall time (\Cref{table:ml_disciplines}). The focus is on one-vs.-one matches and top competitors optimize the time for a certain number of items, rather than optimizing the number of items per time. 

\unnumsubsection{Speed-memory.com}{}
The speed-memory.com website and tournaments were organized by Ram\'on Campayo during the $2000$s and $2010$s \autocite{speed_memory_records}. These competitions featured the most rapid memory disciplines to date: the one and four second memorization of decimal and binary digits, providing us with interesting data on short memory tasks. 


\section{Dramatic increase in performance}
\begin{figure}[!htb]
	\centering
	\includegraphics[width=0.5\linewidth]{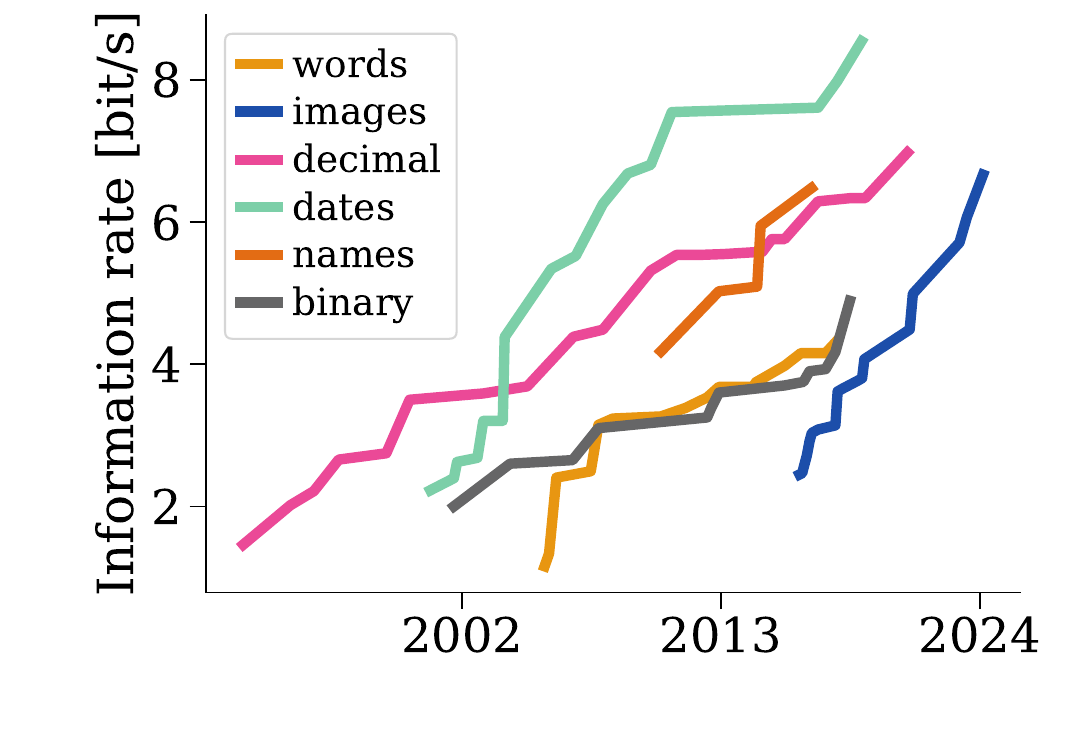}
	\caption{Record development of the classical five-minute disciplines.}
	\label{fig:comparison_all_disciplines}
\end{figure}

Nowadays, tens of thousands of digits are being recalled in $\pi$ competitions. Modern records include memorizing $35$ packs of cards or an over $3000$-digit decimal number in one hour, a sequence of $145$ words in $5$ min, the names of a class of $30$ students in under $30$ s, a shuffled deck of $52$ playing cards in around $12$ s or a $50$-digit binary number in one second.
Calculating information rates of records allows us to visualize and compare the record development of, for instance, all classical five-minute disciplines (\Cref{fig:comparison_all_disciplines}), revealing a striking overall trend. For example, scores in memorizing decimal digits have improved in an almost linear fashion.  

The underlying reasons for the increase are multifaceted and similar to other sports: evolution and disruption of techniques, increased competition and professionalization. 
The digital era has elevated scores. It has become much easier to train the elaborate encoding systems or generate random training sets for the disciplines. Competitors save time by pressing keys instead of flipping pages and scribbling on paper under pressure.
A last effect might be purely psychological: knowing what scores are possible will influence competitors' aims. Unaware of human possibilities, memorizing a $50$-digit decimal number in five minutes was impressive in the early years. Nowadays, knowing that the world record is above $600$, even children and seniors (age $>60$) will try to memorize a few hundred \autocite{WMSC2024}. The latter competitors challenge the common believe that memory abilities are in constant decay after reaching adulthood.

\unnumsubsection{Records will probably improve}{}

Memory sports have gained a wide following over the past three decades and the above factors will continue to contribute to the evolution of records. Despite the enormous increase in scores and competitions, there are still many examples of memory athletes reaching the top of competitive memory sports after only a few years of training, often starting beyond their youth. For instance, Alex Mullen, Andrea Muzii and Enrico Marraffa all became IAM world champions within their first two years of competing.
This likely involves daily training of several hours. Enrico Marraffa, who became world champion in $2024$, stated that he practiced five months roughly three hours a day \autocite{enricomarraffa}. 
Nevertheless, the overall amount of training required is still below Olympic champions, top musicians or professional chess players, where it would be exceptional to reach a world-class level after missing out on the $10$ years of practicing as a child. Therefore, we conjecture that at least some of the records in this document will continue to evolve.


\unnumsubsection{Savants}{}
One might wonder whether there are not a few people with innate ability who could surpass the thresholds presented by memory competitions. As other have noted \autocite{meister}, the world memory champions have all credited training rather than talent. This indicates that innate memorizers are either worse than trained athletes, bound to specific tasks or unwilling to compete \autocite{memorableoutliers}. 

There has been the notable case of Daniel Tammet, who reached global fame with his books \autocite{bornonablueday, thinkinginnumbers}. He describes his memory and mental calculation abilities to originate from a rare combination of synaesthesia and Asperger's syndrome, which became subject of scientific study \autocite{danieltammetstudy}. In turn, Joshua Foer provides evidence in his book ``Moonwalking with Einstein'' \autocite{moonwalkingwitheinstein} that Tammet is using mnemonic strategies. The evidence includes his previous participation in world championships placing $10$th in $1999$ and $4$th in $2000$, a deleted ``about'' webpage, which includes a description of the use of mnemonic strategies and many more. However, Foer writes ``my theory, [...] would be very difficult to prove'' \autocite[p.232]{moonwalkingwitheinstein}. In retrospect, we can at least add the following argument: his performances at the world championships $1999$ and $2000$ are in line with the scores at the time. Modern performances are about three times faster. If it was an innate ability, it seems fairly unlikely that by chance its speed matches the standard of memory competitions at the time.
But didn't Daniel Tammet memorize $22514$ digits of $\pi$? Recently, Susanne Hippauf, a policewoman, broke the German record with $18026$ digits \autocite{pi, hippauf}. If impressive quantities can be memorized given the motivation and technique, when are memory abilities qualifying for labels such as savant or exceptional? The evolution of mnemonic performances creates the impression that acts of extraordinary memory are instead acts of extraordinary motivation.

\section{Mnemonic techniques} \label{sec:mnemonic_techniques}
Essentially, three strategies are used in memory competitions (\Cref{fig:schema_and_reading_times} \textbf{A}) \autocite{artofmemorywiki}:
\begin{enumerate}
	\item \label{strategy_encoding}
	\textit{Encoding} Competitors transform any information into a more concrete experience, object or familiar entity. For instance, an unknown international name might remind the competitor of a word in their native language. 
	
	The more of the tasks structure has already been committed to memory previously, the easier memorization becomes \autocite{domain_expertise_chess, domain_expertise_wiktor}. For example, names in the native language are easier for competitors.
	\item 
	\textit{Associations/Links} \label{strategy_associate} If there are only a few entities to connect, competitors form a story or creative link among them. For example, in the classical IAM images format, where one needs to memorize rows of five images, competitors form stories linking the items. Or in historic dates, where the objective is to memorize the years of fictional historic events, people associate an object or person, which they use to encode the number, with the event.
	\item 
	\textit{Memory Palace} If the task involves memorizing sequential information, competitors link the result of the first two principles with mentally imagined locations of a certain predefined order. For instance, these could be places at home or the way to work. In recall, competitors traverse the same mental route and decode the stories they encounter.
\end{enumerate}
Depending on the disciplines and personal strategy, the three core concepts can be featured more or less prominently. In \Cref{sec:techniques_and_records}, we explain in detail for all disciplines how these concepts come into play and what the current best recorded performances are.

Encoding may occur spontaneously as a creative process, such as for unknown names or words, or may be completely committed to long-term memory if there is a limited set of combinations. In particular, in the disciplines based on digits and cards, competitors have invented elaborate systems to form a map between all possible scenarios and objects or people. In this document, we will refer to the result of this encoding process as an ``mnemonic image'' even though other senses might be involved \autocite{mnemonic_image}. For example, all two-digit decimal numbers could be mapped to mnemonic images, which are then memorized instead. Approaches utilizing modular mnemonic images such as person-action-object (PAO) are also highly successful (\Cref{subsec:digits}).

\section{General observations}

\unnumsubsection{Human beings are no machines}{subsec:no_machines}
In light of the computer-like memory world records, it is easy to forget that competitors are prone to errors, which causes the number of attempts and grading systems to have significant influence on the outcomes. Fewer attempts result in fewer samples from an mnemonist's performance distribution. Additionally, competitors are risk-averse and might, for instance, attempt a score sufficient to win rather than an exceptional score. The grading system of errors plays an equally crucial role. Counting only perfect performances as records, such as in ML, classical auditive digits and $\pi$ competitions, leads to significantly lower scores.

For example, in auditive digits, the objective is to memorize a sequence of decimal digits, read aloud at one digit/s. The score is the number of digits counted until the first error. The current world record of Lance Tschirhart is $456$, over seven minutes of perfect memorization. Notably, this is close to the current average speed of $0.95$ digit/s of the one hour digits marathon. Confronted with a specific speed, humans fail quickly, whereas at varying speed and with revisions they can keep up the same average for much longer. 

\begin{figure}[!htb]
	\centering
	\resizebox{\textwidth}{!}{
		\begin{tikzpicture}
			
			\node[anchor=west, font=\bfseries\large\rmfamily] at (0,9.6) {A}; 
			\node[anchor=center, font=\large\rmfamily, fill=gray!10, rounded corners] at (7.5,9.6) {Subtasks underlying the memory palace};
			
			\node[anchor=center, font=\normalsize\rmfamily] at (2.5,9) {Reading};
			\node[rounded corners, fill=gray!10, minimum width=4cm, minimum height=3cm] at (2.5,7.25) {}; 
			\node[anchor=center] at (3.25,7.85) {\includegraphics[width=0.14\textwidth]{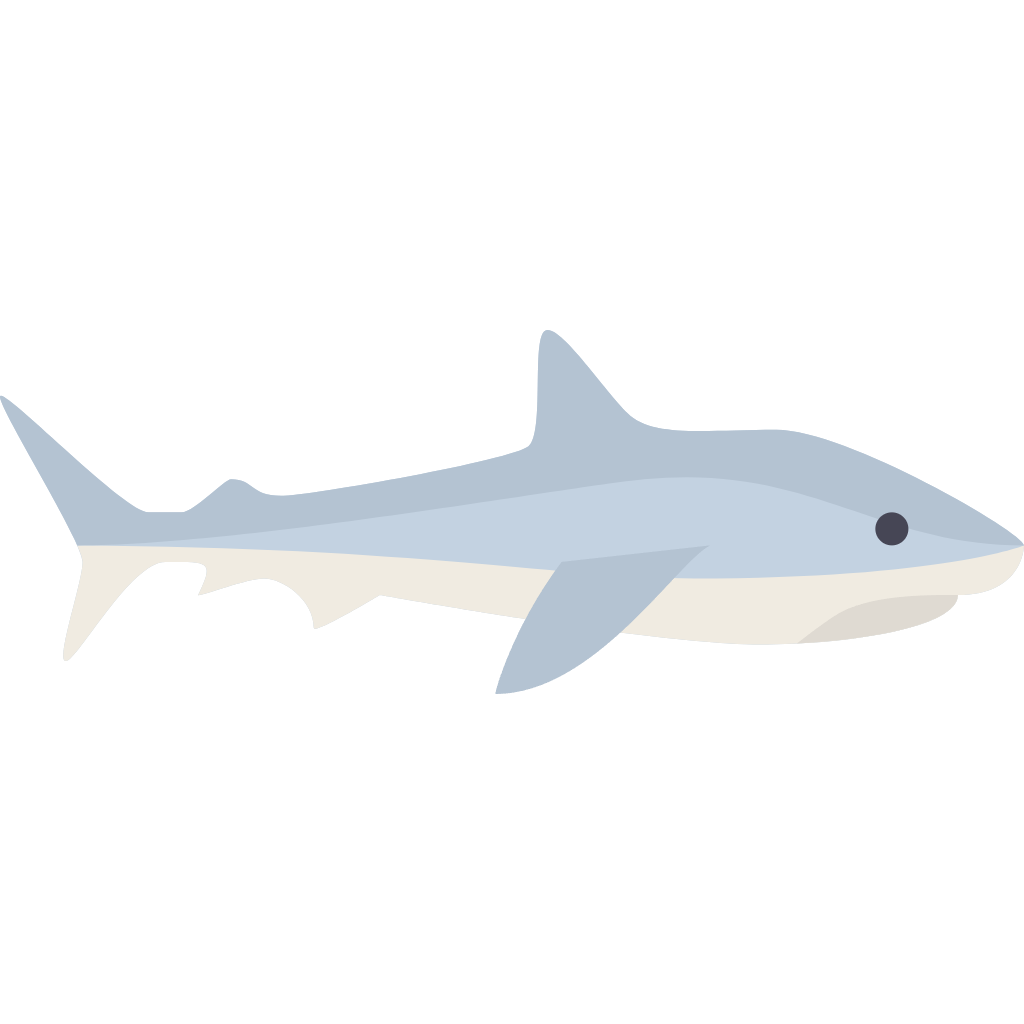}};
			\node[anchor=center, font=\Large\rmfamily, align=center] at (1.5,7.85) {647 \textcolor{gray!70}{=}};
			\node[anchor=center] at (3.25,6.65) {\includegraphics[width=0.07\textwidth]{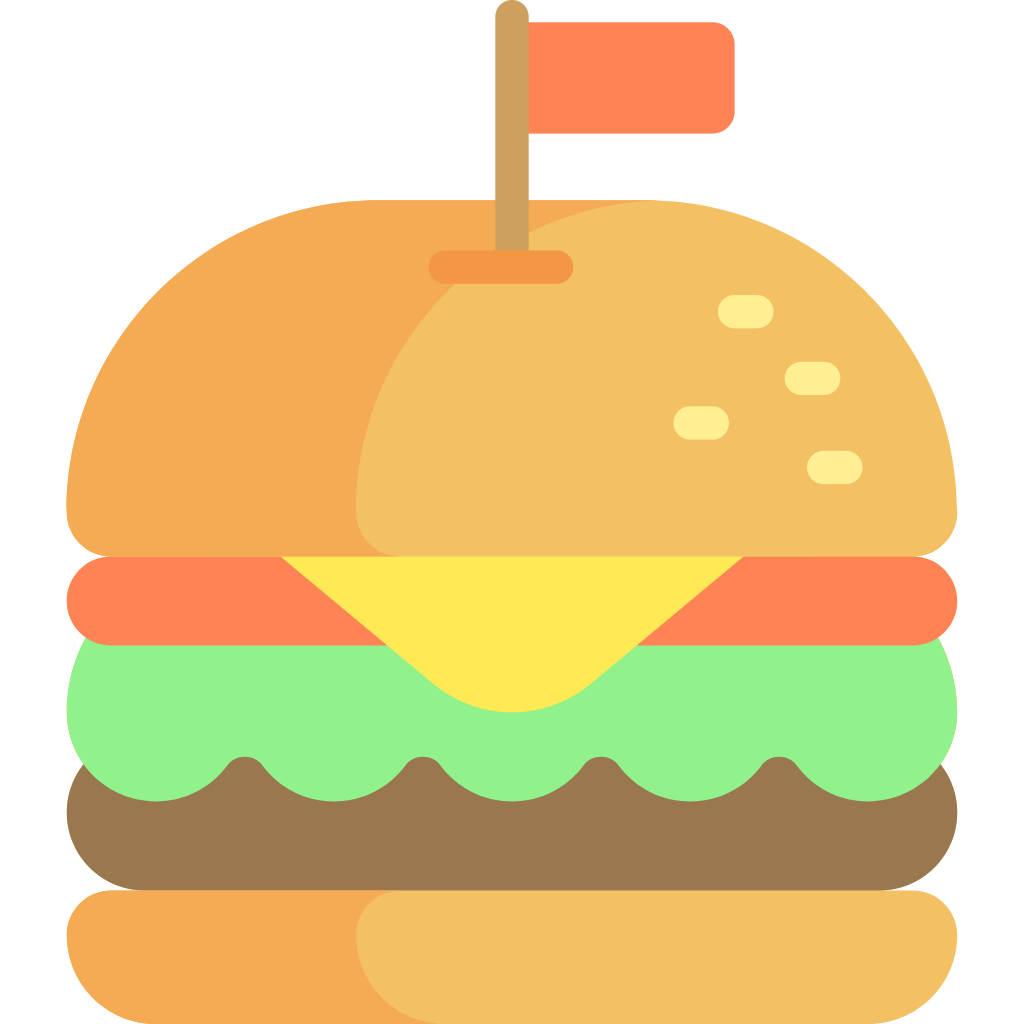}};
			\node[anchor=center, font=\Large\rmfamily, align=center] at (1.5,6.55) {639 \textcolor{gray!70}{=}};
			
			\draw [rounded corners,fill=gray!20, draw=none] (4.75,6)--(4.75,8.5)--(5.25,7.25)--cycle;
			
			\node[anchor=center, font=\normalsize\rmfamily] at (7.5,9) {Associations};
			\node[rounded corners, fill=gray!10, minimum width=4cm, minimum height=3cm] at (7.5,7.25) {};
			\node[anchor=center] at (8.7,7.15) {\includegraphics[width=0.04\textwidth, angle=-45]{Figures/burger.png}};
			\node[anchor=center] at (7.35,7.25) {\includegraphics[width=0.19\textwidth]{Figures/shark.png}};
			\node[anchor=center] at (8.7,7.15) {\includegraphics[width=0.04\textwidth, angle=-45]{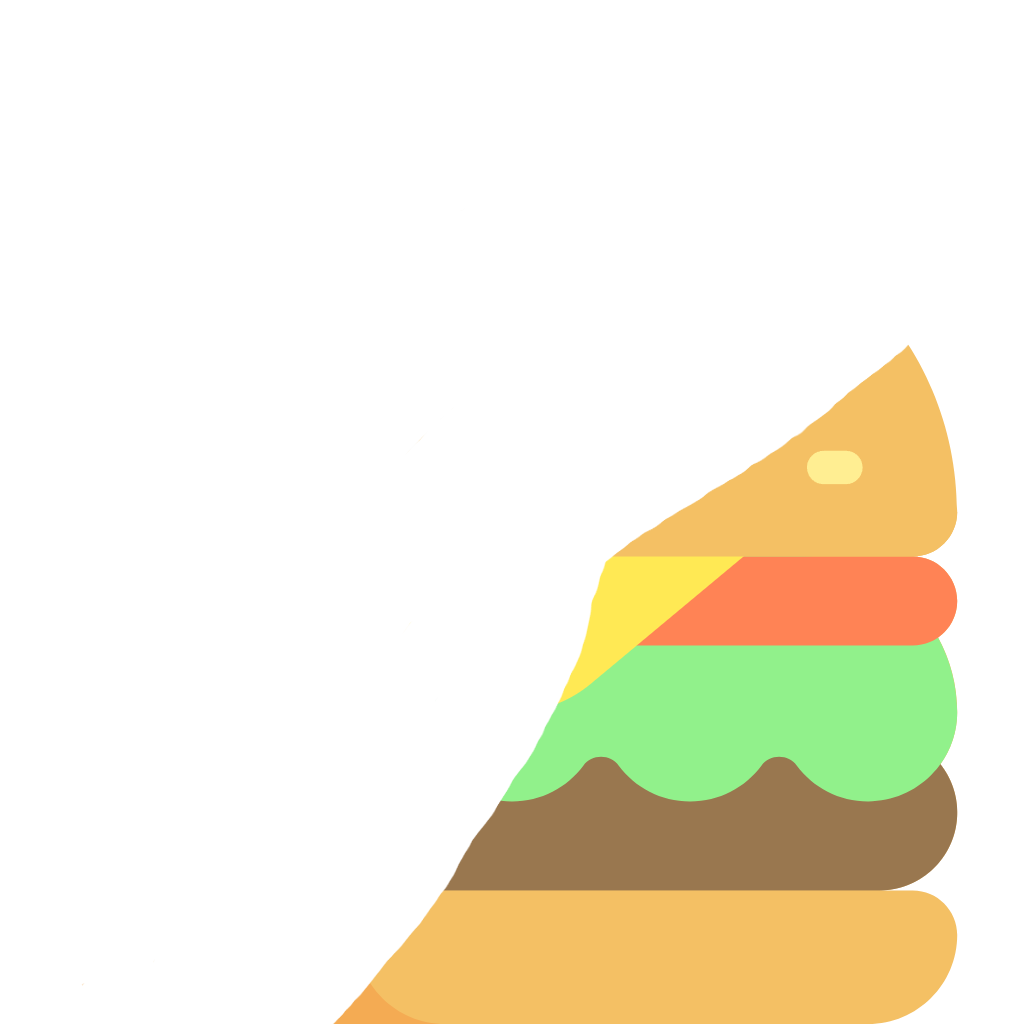}};

			\draw [rounded corners,fill=gray!20, draw=none] (9.75,6)--(9.75,8.5)--(10.25,7.25)--cycle;
			
			\node[anchor=center, font=\normalsize\rmfamily] at (12.5,9) {Navigation};
			\node[rounded corners, fill=gray!10, minimum width=4cm, minimum height=3cm] at (12.5,7.25) {};
			
			\node[anchor=center] at (11.8,7.32) {\includegraphics[width=0.12\textwidth, angle=-90]{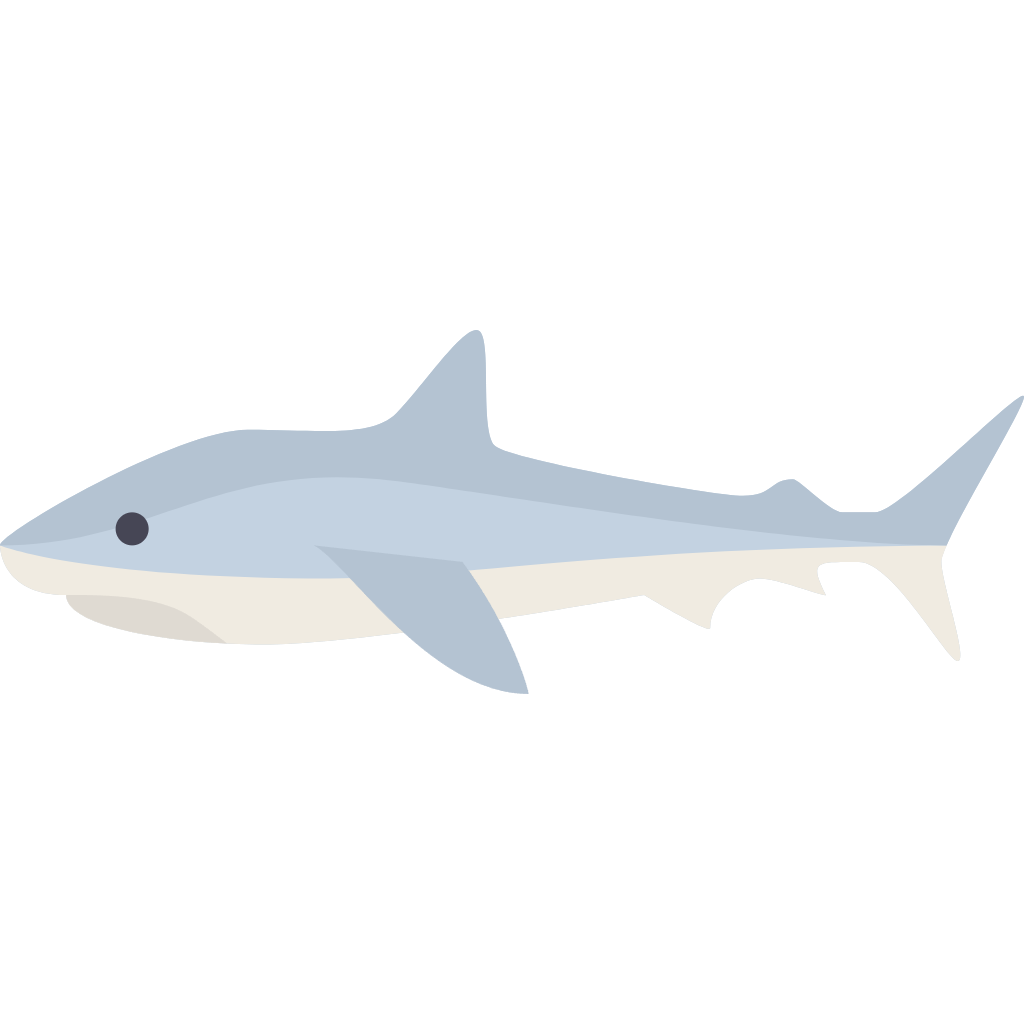}};

			\node[anchor=center] at (13.5,8.25) {\includegraphics[width=0.05\textwidth]{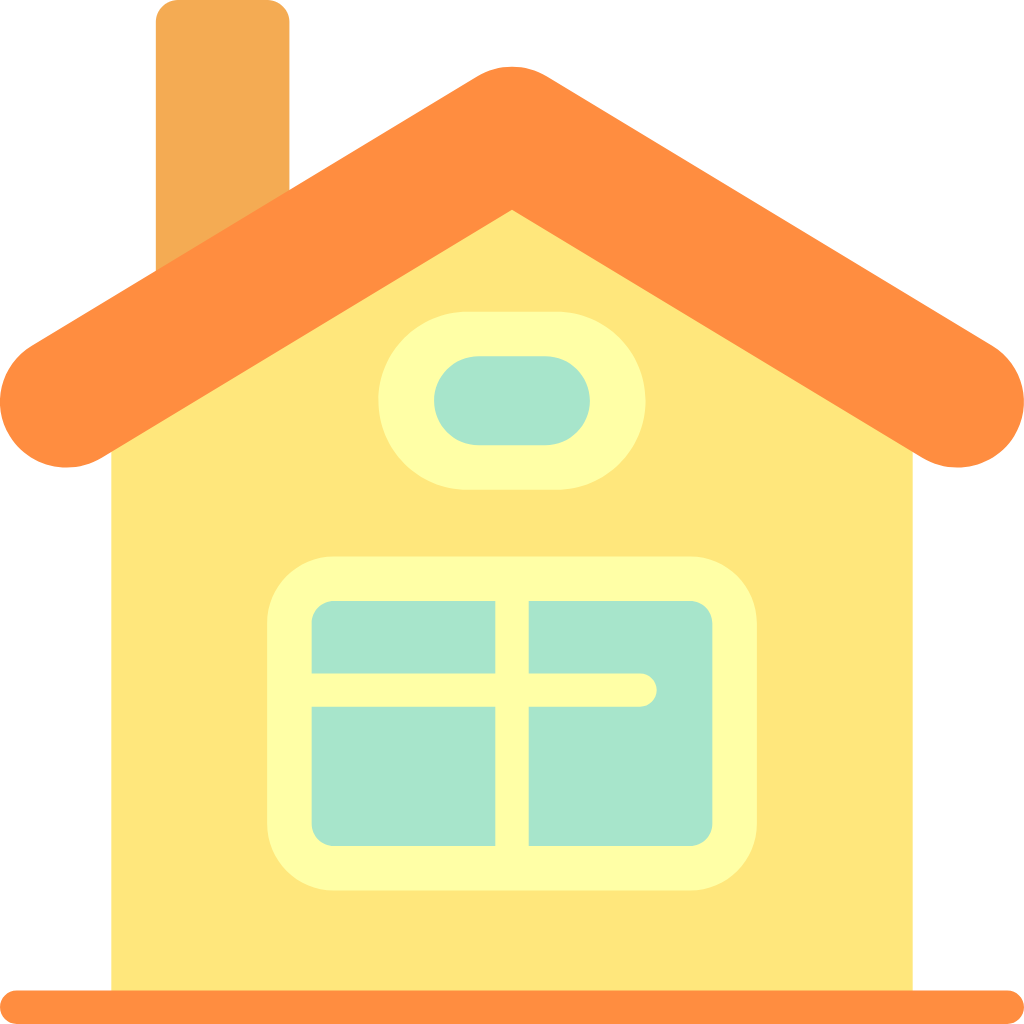}};
			\node[anchor=center] at (11.3,7.05) {\includegraphics[width=0.08\textwidth]{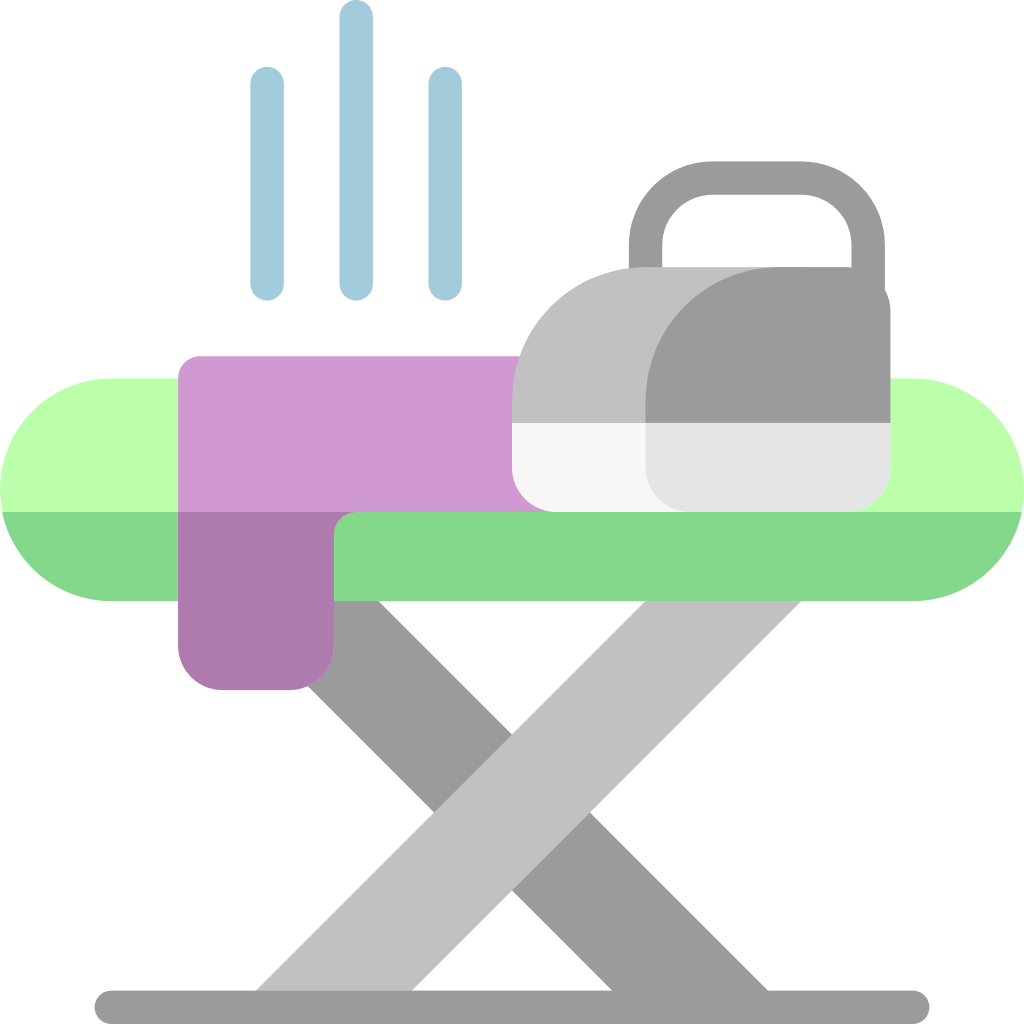}};
			\node[anchor=center] at (13.5,6.25) {\includegraphics[width=0.04\textwidth]{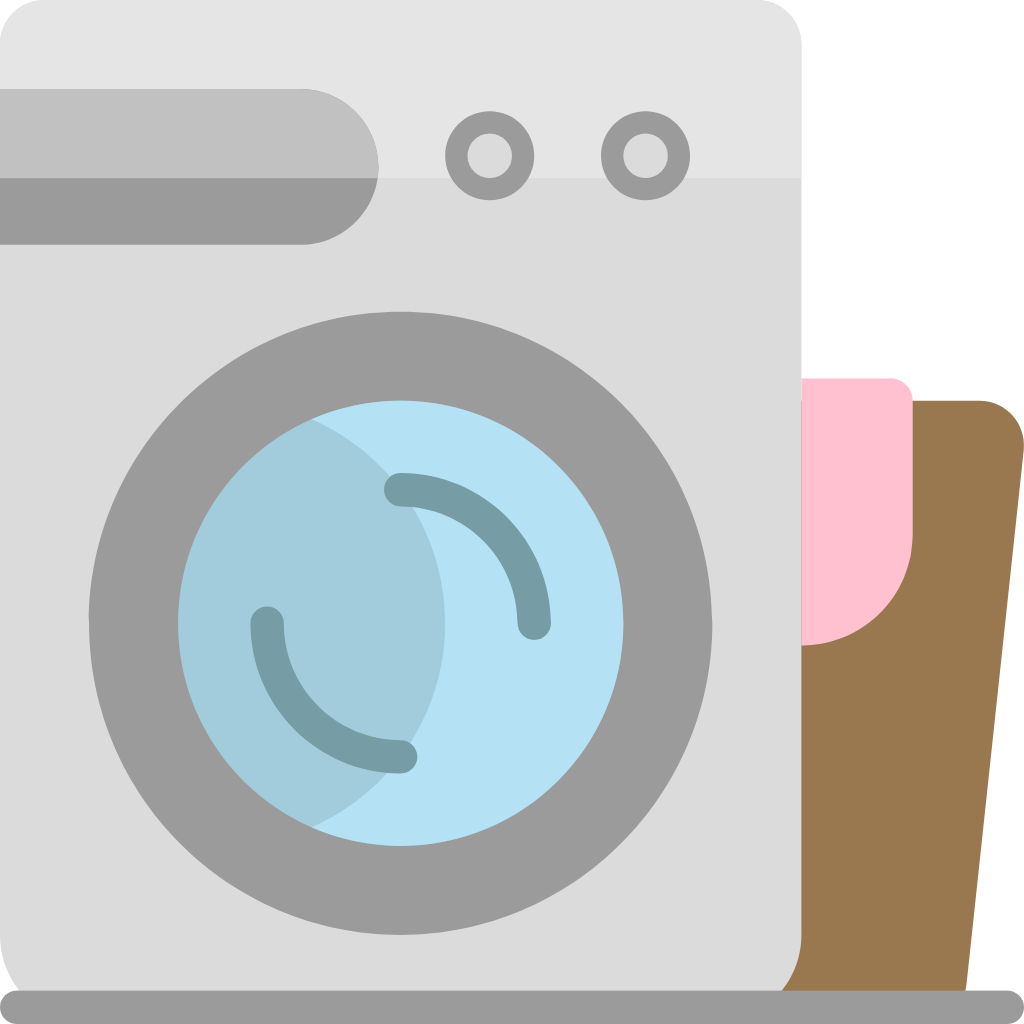}};
			
			\node[anchor=center] at (11.73,8.18) {\includegraphics[width=0.025\textwidth, angle=-45]{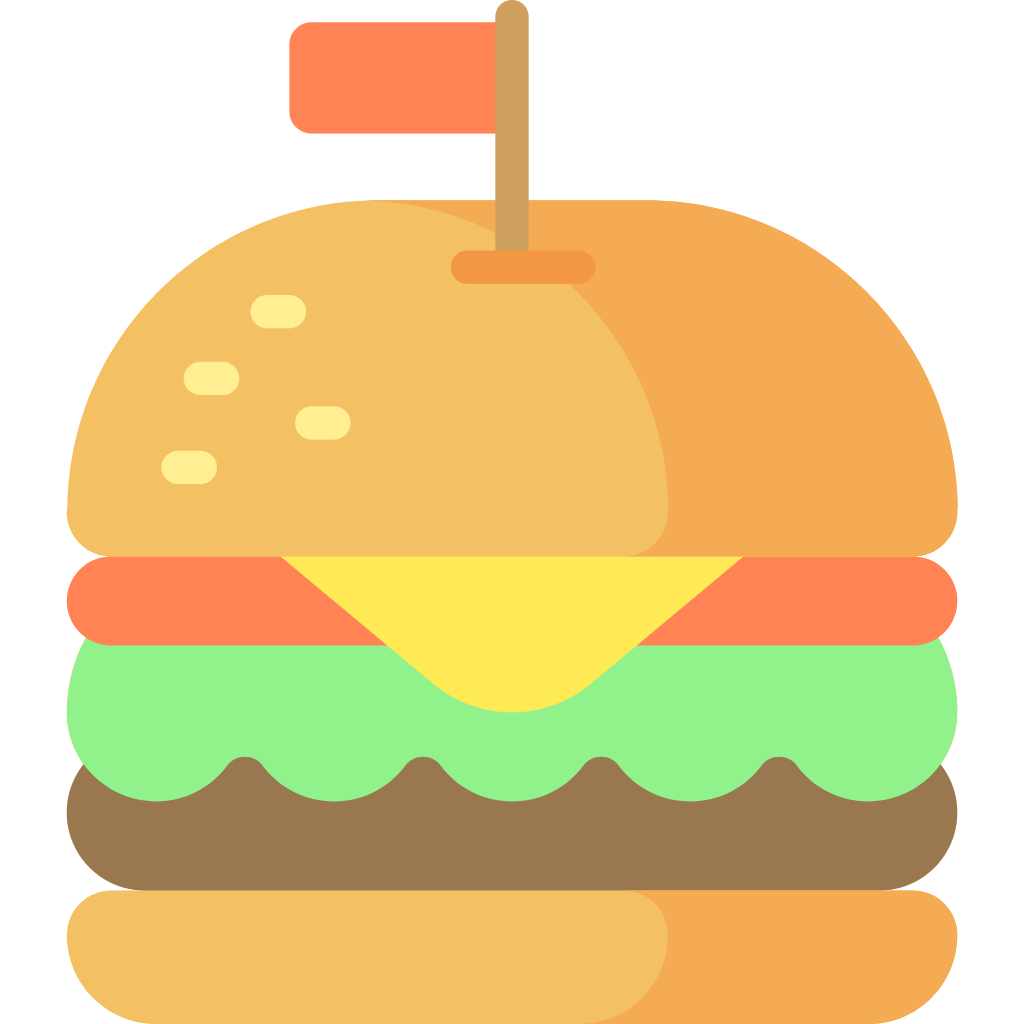}};
			\node[anchor=center] at (11.8,7.32) {\includegraphics[width=0.12\textwidth, angle=-90]{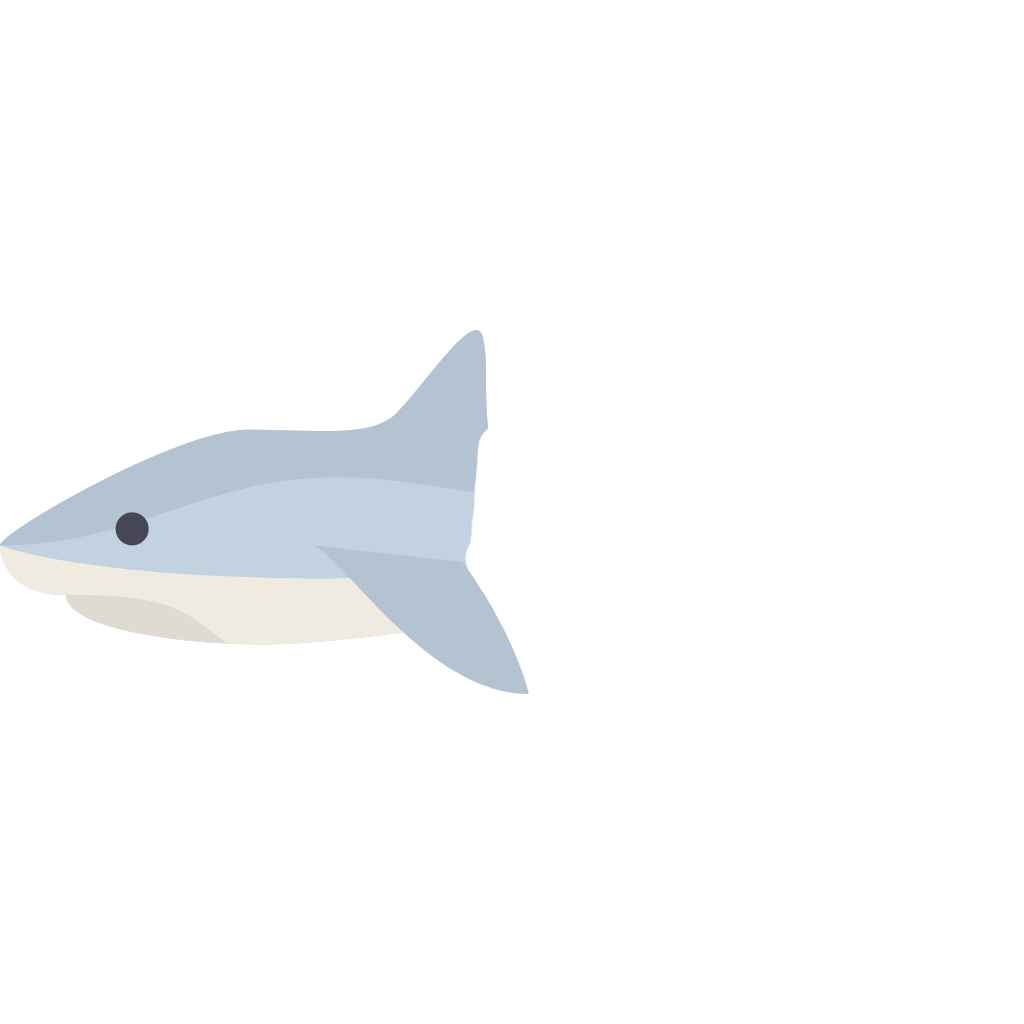}};
			\node[anchor=center] at (11.73,8.18) {\includegraphics[width=0.025\textwidth, angle=-45]{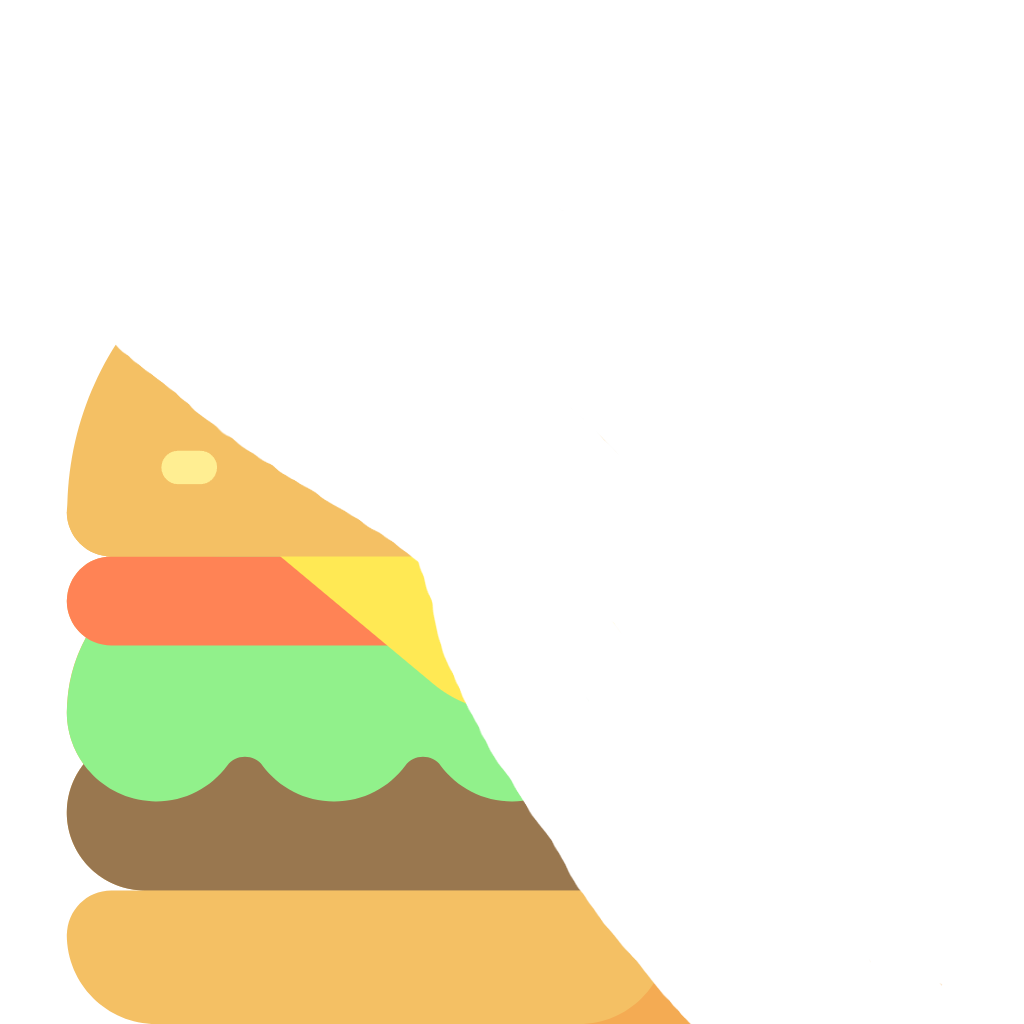}};
			\fill[gray!70] (12.35,7.75) circle (2pt);
			\fill[gray!70] (12.6,7.9) circle (2pt);
			\fill[gray!70] (12.85,8.05) circle (2pt);
			\fill[gray!70] (12.35,6.75) circle (2pt);
			\fill[gray!70] (12.6,6.6) circle (2pt);
			\fill[gray!70] (12.85,6.45) circle (2pt);

			\node[anchor=west, font=\bfseries\large\rmfamily] at (0,5) {B}; 
			\node[anchor=center, font=\large\rmfamily, fill=gray!10, rounded corners] at (7.5,5) {Performances and reading times of selected memory athletes};
			
			\node[anchor=north west] at (0, 4.25) {
				\def\spacingtabular{12pt}
				\footnotesize
				\begin{tabular}{C{11mm} C{22mm} C{10mm} C{31mm} C{11mm} C{20mm} C{9mm}}
					& Performance (P) & P $[\frac{\text{bit}}{s}]$ & Reading time (RT) & RT $[\frac{\text{bit}}{s}]$ & Difference ($\Delta$) &  $\Delta \, [\frac{\text{bit}}{s}]$ \\[4pt]
					\hline & & & & & & \\[-1ex]
					Alex Mullen & \makecell{ML cards \\ $11.89$ s $52/52$\\ \text{see methods}} & $18.97$ & $200$ ms estimate $\frac{52}{2} = 26$ items $\simeq 5.2$ s & $\approx 43$ & $6.7$ s & $\approx 34$ \\[\spacingtabular]
					& & & Self-reported $\approx 9$ s & $\approx 25$ & $2.9$ s & $\approx 78$ \\[\spacingtabular]
					Andrea Muzii & \makecell{ML digits \\ $9.75$ s $80/80$\\ \autocite{andreadigitsyoutube}} & $ 27.25$ & $200$ ms estimate $\lceil \frac{80}{3} \rceil = 27$ items $\simeq 5.4$ s & $\approx 49$ & $4.4$ s & $\approx 60$ \\[\spacingtabular]
					& & & Self-reported $\approx 8.25$ s & $\approx 32$ & $1.5$ s & $\approx 177$ \\[\spacingtabular]
					\makecell{Don\\ Michael\\ Vickers} & \makecell{ML words\\ $12.77$ s $45/50$\\ \autocite{donwordsyoutube}} & $42.04$ & \makecell{$200$ ms estimate\\ $45$ items $\simeq 9$ s} & $\approx 60$ & $3.8$ s & $\approx 141$ \\[\spacingtabular]
					& & & Self-reported $\approx 9$ s & $\approx 60$ & $3.8$ s & $\approx 141$ \\
				\end{tabular} 
			};
		\end{tikzpicture}
	}
	\caption{\textbf{A} Schematic of the process of using a memory palace. First, the input, in this case numbers, are transformed into mnemonic images. Second, associations are formed between the mnemonic images and also the position of the mental walk. Once a position is occupied, competitors navigate mentally to the next salient location. In most cases, the mental walk is along a real-world location.\\
		\textbf{B} Performances, estimated minimal and self-reported reading times, differences $\Delta$ between the reading and memorization time and the associated information rates for selected memory athletes. To determine the length of the reading phase, we asked the competitors and estimated the minimum based on $200$ ms per item \autocite{singlewordprocessing}. Even the latter conservative estimates constitute a large fraction of the memorization time. If reading and the formation of associations were entirely sequential, $\Delta$ would be the association/memorization time corresponding to the speed in the right-most column. In ML cards the task is to memorize the order of $52$ cards in a shuffled deck, in ML digits an $80$-digit decimal number and in ML words a sequence of $50$ words. Andrea Muzii is using a three-digit system, converting three-digit decimal numbers into one mnemonic image, so there are $27$ items to be perceived \autocite{andreathreedigits}. Alex Mullen achieved the time with a system encoding two cards in one mnemonic image, so $26$ items need to be perceived \autocite{alexhalftwocard}.}
	\label{fig:schema_and_reading_times}
\end{figure}

\unnumsubsection{The limit is reading rather than memorizing}{subsubsec:limit_is_reading}
Using a memory palace involves three different processes: reading, associating and navigating (\Cref{fig:schema_and_reading_times} \textbf{A}). For example, a competitor in the decimal digits discipline first maps the visual input, the digits, onto specific prememorized mnemonic images. This process is comparable to standard reading as, indeed, the most popular strategy to learn the map is to assign letters to the digits $0-9$ \autocite{major_system}. The resulting images are associated to salient positions of the mental walk through short creative stories. Once a location is occupied, the competitor mentally navigates to the next location in the memory palace. Among these three processes, it seems most natural to equate ``memorization'' with the process of forming associations. Surprisingly, rather than this association step, it seems that the process of reading is a substantial bottleneck.

To understand the time spent on the three processes, we asked top competitors in ML cards, digits and words for the time they require to read the items of the tasks (\Cref{fig:schema_and_reading_times} \textbf{B}). The self-reported reading times could be fairly accurate, as competitors train extensively over years to minimize total time. In ML words, reading is comparable to everyone's understanding of the term, whereas in ML cards and digits, reading refers to the aforementioned transformation into mnemonic images. 

Furthermore, we calculated a realistic minimum of the reading time based on research on single-word reading, which suggests that $200$ milliseconds elapse due to visual processing before the visual input becomes available to the mnemonists' minds \autocite{singlewordprocessing}. We received values in line with the self-reported times and the estimate of $50$ bit/s for general reading \autocite{human_information_channel}.

Self-reported times and the estimated minima indicate that reading requires most of the memorization time. If mental associations were formed solely during the difference $\Delta$ between the memorization time and the reading time, processing rates in memory would be significantly higher than indicated by the task itself. 

A recent study revealed the conundrum of human cognition: whereas our sensory systems are able to process enormous quantities of information, in the range of $10^9$ bit/s for the retina, most human activities, from playing Tetris over memorization to typing, exhibit only about $10$ bit/s of processing \autocite{meister}. The prevalence of parallel processing in sensory systems and serial processing in high-level cognition was identified as one of the major contributing factors.
The significant fraction of time spent on reading in rapid memorization tasks, performed at up to $42$ bit/s, illustrates that either there is more high-level parallelization than suggested or that, excluding perceptual time, the human brain can be way faster. If reading, associating and navigating were parallel, this would illustrate that even though humans seem to perform one action at the time, highest human processing speed can be supported by an underlying parallelization of the involved brain regions. If the processes were instead sequential, the narrow difference between memorization and reading time would imply that the actual process of memorization reaches around $150$ bit/s. Equally, any combination of the two options would explain part of the conundrum. It would be interesting to understand the speed of different cognitive subprocesses and how much parallelization exists during the use of memory palaces by top competitors.

\begin{figure}[!htb]
	\centering
	\centering
	\resizebox{\textwidth}{!}{
		\begin{tikzpicture}
			\fill[white] (14.5,0) circle (2pt); 
			\node[anchor=west, font=\bfseries\large\rmfamily] at (0,0) {A}; 
			\node[anchor=center, font=\large\rmfamily, fill=gray!10, rounded corners] at (7.5,0) {Official records reveal power-law decrease of rates};
			
			\node[anchor=center, font=\normalsize\rmfamily] at (3,-0.6) {All records};
			\node[anchor=north west] at (0,-0.8) {\includegraphics[width=0.335\textwidth]{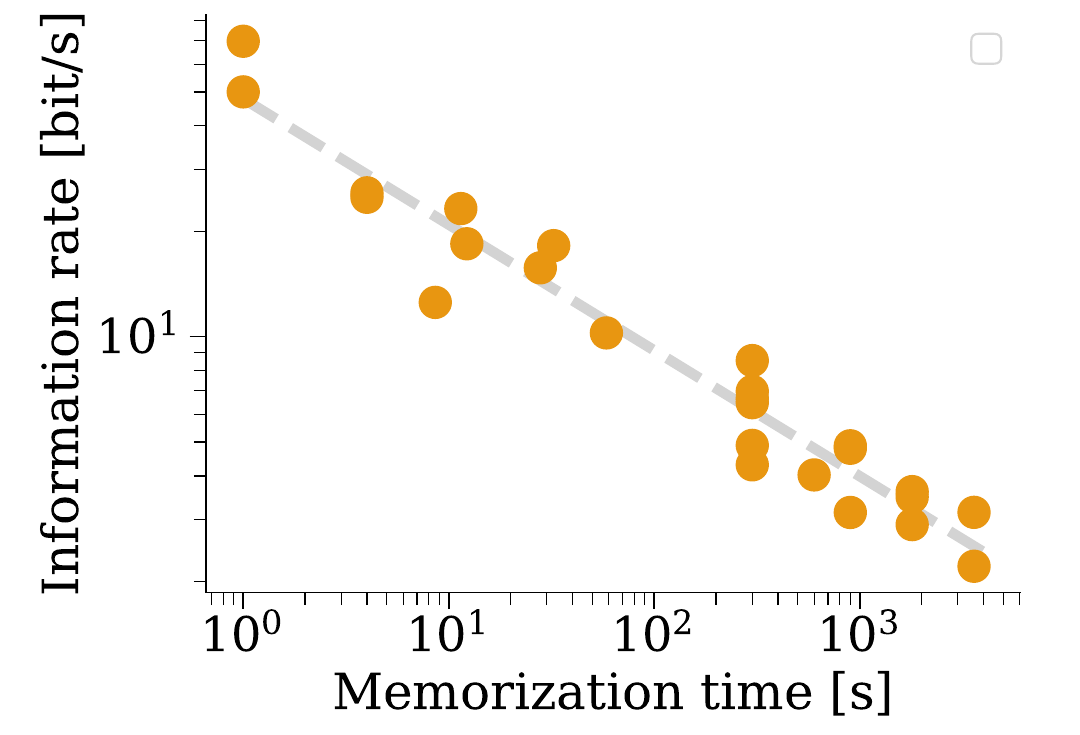}};
			\node[anchor=center, font=\normalsize\rmfamily] at (10,-0.6) {Subcategories};
			\node[anchor=north west] at (5,-0.8) {\includegraphics[width=0.335\textwidth]{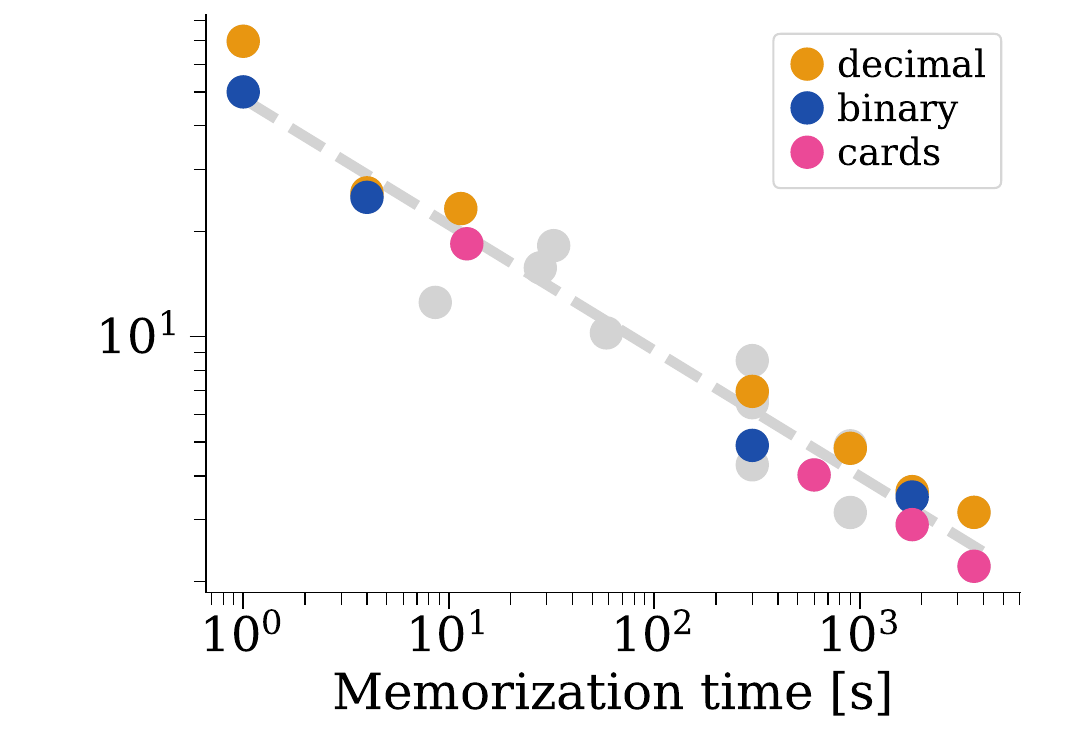}};
			\node[anchor=north west] at (10,-0.8) {\includegraphics[width=0.335\textwidth]{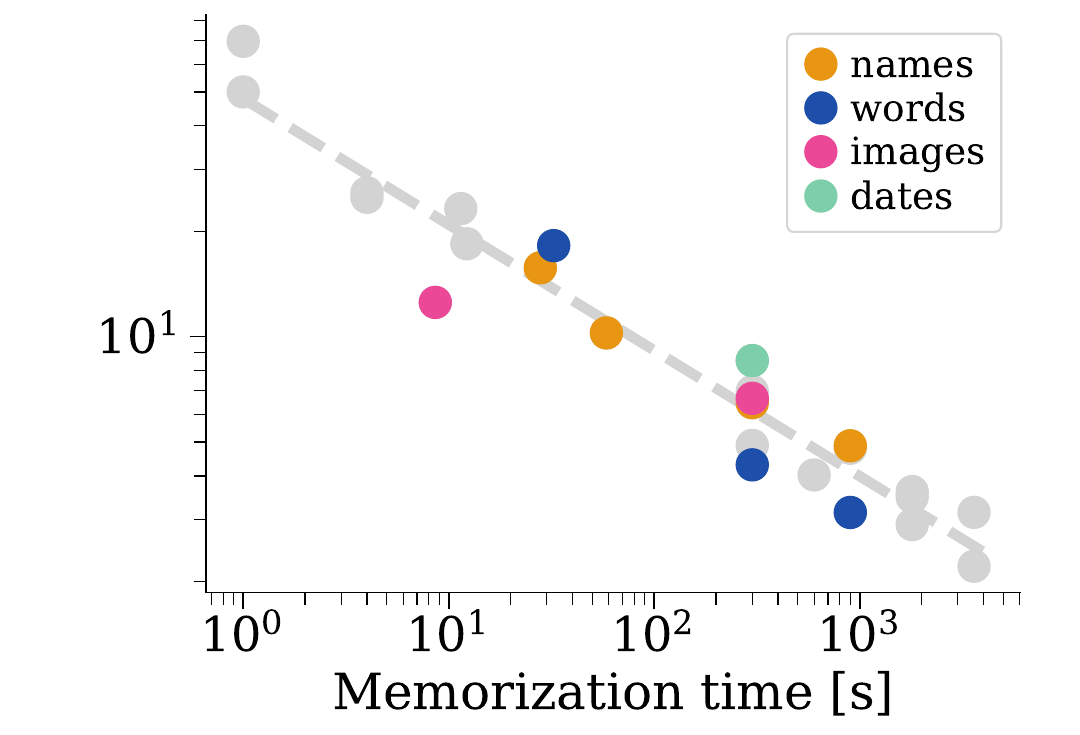}};
			\node[anchor=center, font=\normalsize\rmfamily] at (3,-5) {All records};
			\node[anchor=north west] at (0,-5.4) {\includegraphics[width=0.335\textwidth]{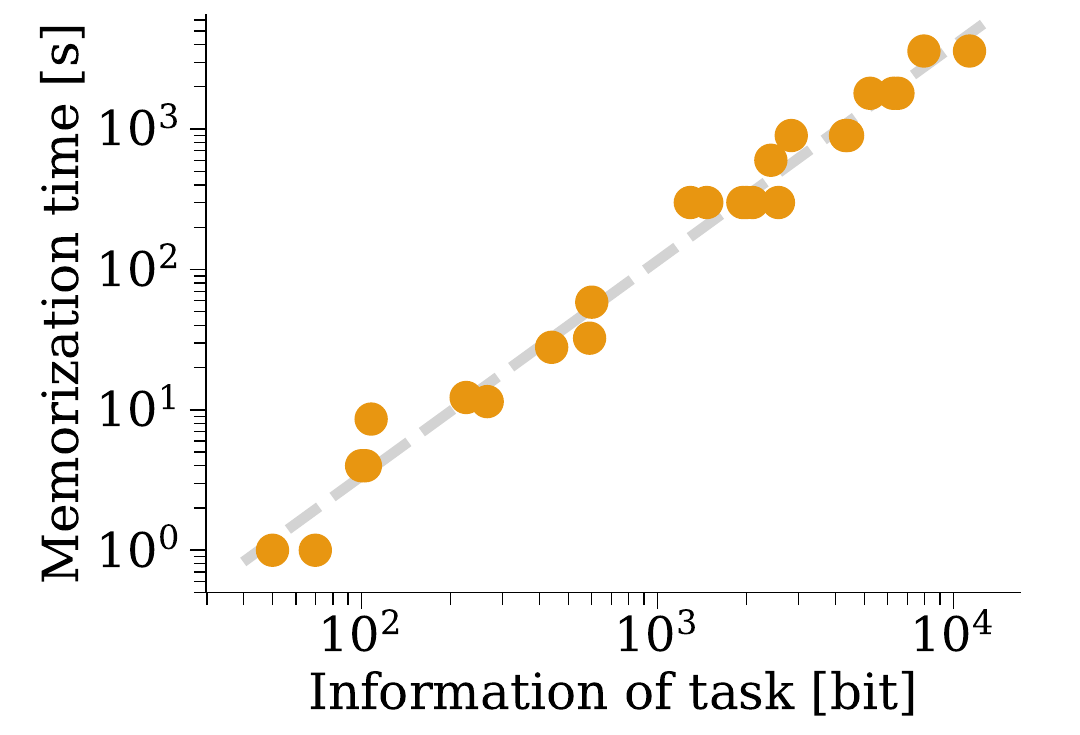}};
			
			\node[anchor=west, font=\bfseries\large\rmfamily] at (5.25,-5) {B}; 
			\node[anchor=center, font=\large\rmfamily, fill=gray!10, rounded corners] at (10.25,-5) {Model prediction and data for ML digits};
			\node[anchor=north west] at (7.5,-5.4) {\includegraphics[width=0.335\textwidth]{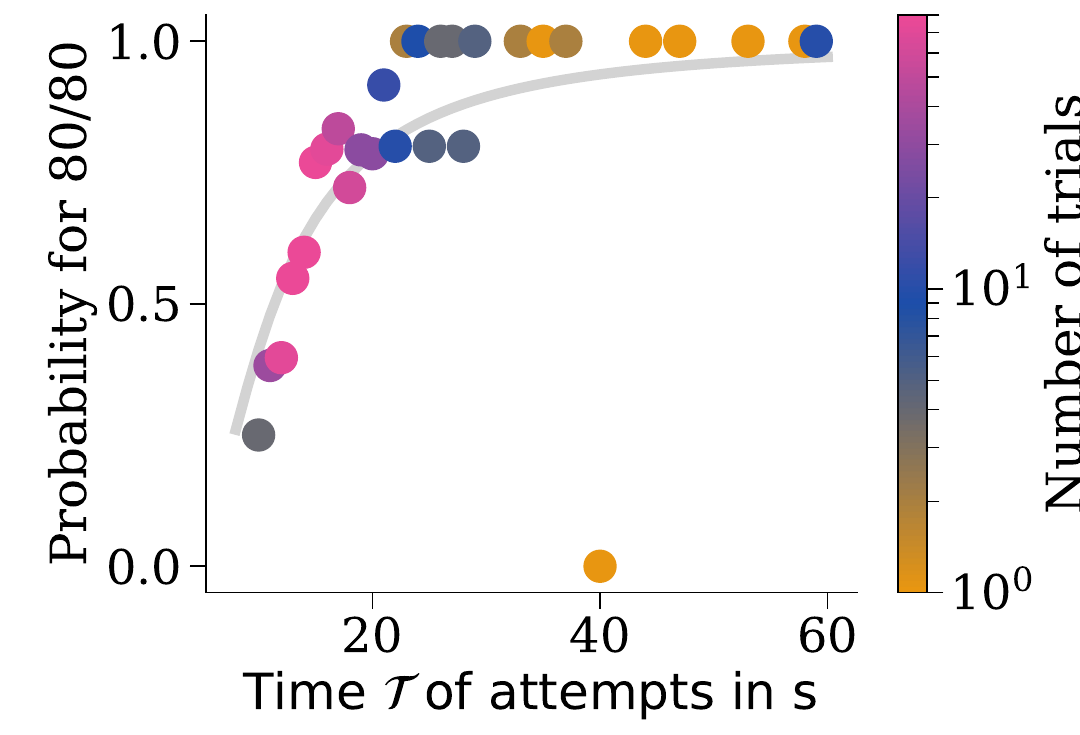}};
		\end{tikzpicture}
	}
	\caption{\textbf{A} The left panel in the top row depicts the information rates of all official memory world records as a function of the duration of the memorization time. The results of $1$ s and $4$ s memorization duration are from speed-memory.com. The remaining results with a memorization time smaller than $60$ s are ML records and the bottom right-hand cloud corresponds to the records from classical competitions. The vertical alignment of parts of the data results from the memorization time being either $5, 15, 30$ or $60$ min at classical competitions. The speed-card record in classical competitions of Shijir-Erdene Bath-Enkh is $12.74$ s. It is reassuring that the score is similar to the ML cards record of $12.25$ s, as this is the only discipline directly comparable between ML and classical competitions. The middle and right panel of the top row show the same records, but grouped by discipline categories.
		The bottom panel shows instead memorization duration as a function of the information of the complete tasks.\\
		\textbf{B} Empirical success probabilities for achieving $80/80$ in ML digits compared to the prediction of the probabilistic model. Shown are data for the success probability of trials, binned in intervals of one second, by competitors Alex Mullen and Andrea Muzii, the official and unofficial ML record holders. Note that around $85\%$ of trials are in the time between $10$ and $20$ s, and larger times are insufficiently sampled. For example, there was only a single attempt in the $40$ s bin. The curve depicts the prediction $\mathbb{P}(A(80, \mathcal{T} - r))$ of the model. We used the parameters of the power law \ref{eq:power-law}, a reading time $r$ of $8$ s (\Cref{fig:schema_and_reading_times} \textbf{B}) and obtained $p = 0.72$ for the free parameter of the model using weighted least squares, with the number of attempts of a certain time as weights.} 
	\label{fig:power_law_and_model}
\end{figure}

\unnumsubsection{Power law of memorization speed depending on time span}{subsec:power_law}
Across memorization tasks, the greater the number of items to be memorized, the longer the time competitors require. The reading time per item is not expected to increase decisively, but more time is required to consolidate the information in memory until the start of the recall phase. It is, therefore, not surprising that the information rates are lower for longer tasks. Intriguingly, when we calculated the minimum information rates necessary to explain the records, we found a power law of information rate over the memorization time (\Cref{fig:power_law_and_model} \textbf{A}). We fitted the function $a T^b$ for two parameters $a,b$ to the data points provided by the official records using least squares based on log quantities. As a function of the duration $T$ of the memorization phase in seconds, the resulting curve of the information rate $R$ is
\begin{equation} \label{eq:power-law}
	R (T) \approx 47.79 \cdot T^{- 0.36} \enspace  \frac{\text{bit}}{s}.
\end{equation}
Note that, compared to standard forgetting power laws \autocite{power-law_memory}, \Cref{eq:power-law} does not describe the proportion of correctly recalled items over time, but rather the achieved information rates over different time spans. The dependence of the memorization time on the information of the complete task is also described by a power law rather than a linear relationship (\Cref{fig:power_law_and_model} \textbf{B}). 
As a function of the complete memorized information $I$ in bit, the memorization time in seconds is described by
\begin{equation*}
	R' (I) \approx 29.1 \cdot 10^{-4} \cdot I^{1.53} \enspace s.
\end{equation*}

 Using a simple probabilistic model, we related the power law to the probability to remain error-free in tasks with a certain number of items. The model is based on the assumption that competitors aim for error-free memorization and that long tasks can be divided into shorter tasks with independent error probabilities. The model predicts that, given enough time, competitors' success-rate for perfect memorization of a certain number of items is high, but rapidly decreases if the memorization time becomes shorter than a certain threshold. This seems to match observations on the error-rates in ML (\Cref{fig:power_law_and_model}).

It is plausible that the power-law decrease in the bit rate with time is a consequence of a larger time investment in revision and in better and more vivid associations to overcome the forgetting curve \autocite{forgettingcurve, forgettingcurve_form}.
During memorization, competitors may read the information only once at a slower pace, or revise twice or even several times in longer disciplines.
As revision strategies of participants vary, the number of attempts on longer disciplines is smaller and we have no statistics on competitors errors, future controlled experiments will uncover the precise principles of the power law. 

We tried to estimate the memorization time of the $\pi$-record using the function $R (t)$. The $\pi$ world record of Suresh Kamar Sharma stands at an impressive $70030$ digits, a multiple of around $20.5$ of the hour world record of $3412$ \autocite{pi}. At first sight, this factor might seem small given the unlimited memorization time. However, in the hour digits discipline errors are tolerated.
Aiming for $200 000$ digits to account for the different grading schemes, the use of $R$ returns roughly $130$ hours. In reality, the required amount of time is thousands of hours \autocite{pi_superior_memory}. Our underestimate illustrates that large time scales and capacities pose additional challenges. The discipline places unique demands on encoding system variability as similar combinations repeat. The physical task of reciting the digits is a marathon. Competitors have a few thousands memory palace loci for competitions \autocite{alexnumberofloci}, but the $\pi$-records have become so large that they need to create additional ones. Interestingly, the unlimited memorization time renders memory palaces less relevant, as competitors can create long stories particularly tailored towards the sequence of digits of $\pi$ through various encoding schemes \autocite{piphilology}.

\begin{figure}[!htb]
	\centering
	\resizebox{\textwidth}{!}{
		\begin{tikzpicture}
			
			\node[anchor=west, font=\bfseries\large\rmfamily] at (0,0) {A}; 
			\node[anchor=center, font=\large\rmfamily, fill=gray!10, rounded corners] at (7.5,0) {Remarkable record development};
			\node[anchor=center, font=\normalsize\rmfamily] at (3,-0.6) {Five minute disciplines};
			\node[anchor=north west] at (0,-0.8) {\includegraphics[width=0.335\textwidth]{Figures/comparison_all_disciplines.pdf}};
			\node[anchor=center, font=\normalsize\rmfamily] at (8,-0.6) {Decimal digits};
			\node[anchor=north west] at (5,-0.8) {\includegraphics[width=0.335\textwidth]{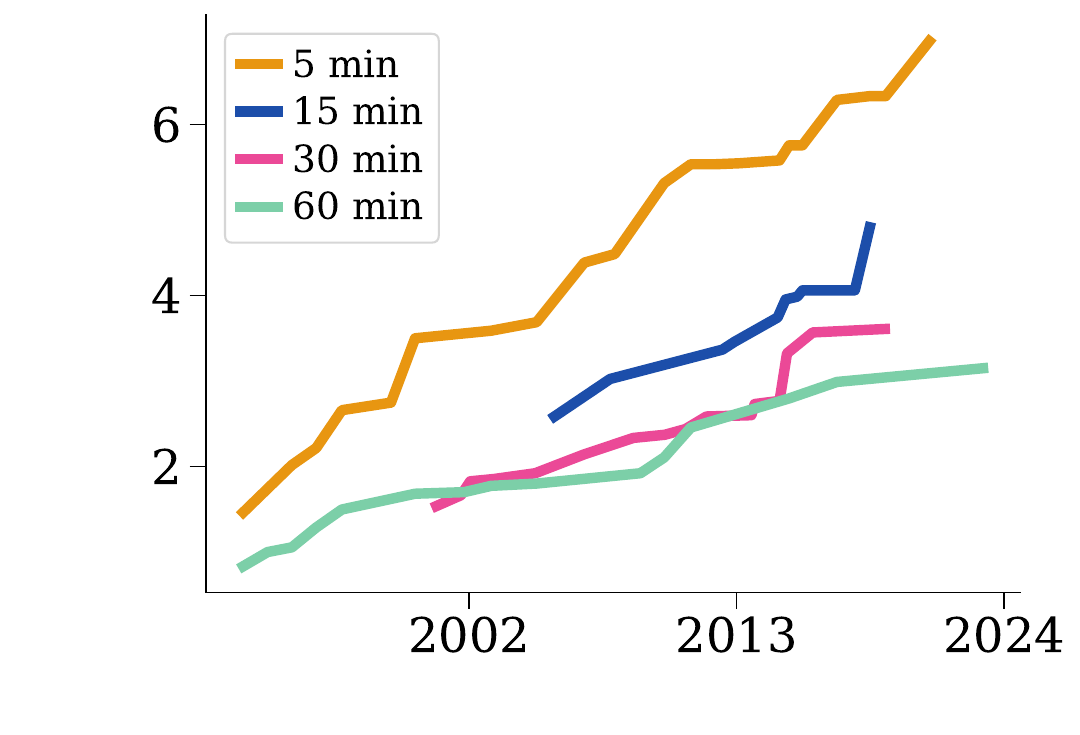}};
			\node[anchor=center, font=\normalsize\rmfamily] at (13,-0.6) {Cards};
			\node[anchor=north west] at (10,-0.8) {\includegraphics[width=0.335\textwidth]{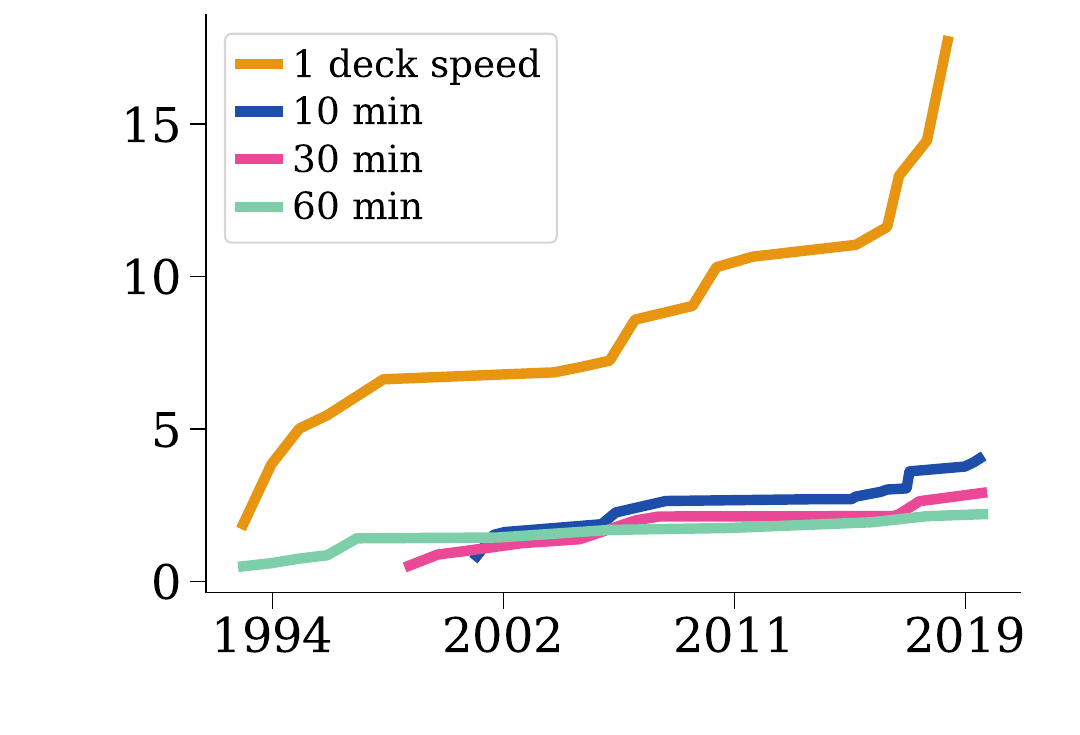}};
			
			\node[anchor=west, font=\bfseries\large\rmfamily] at (0,-4.7) {B}; 
			\node[anchor=center, font=\large\rmfamily, fill=gray!10, rounded corners] at (7.5,-4.7) {Consistent differences between disciplines};
			\node[anchor=north west] at (0,-5.5) {\includegraphics[width=0.335\textwidth]{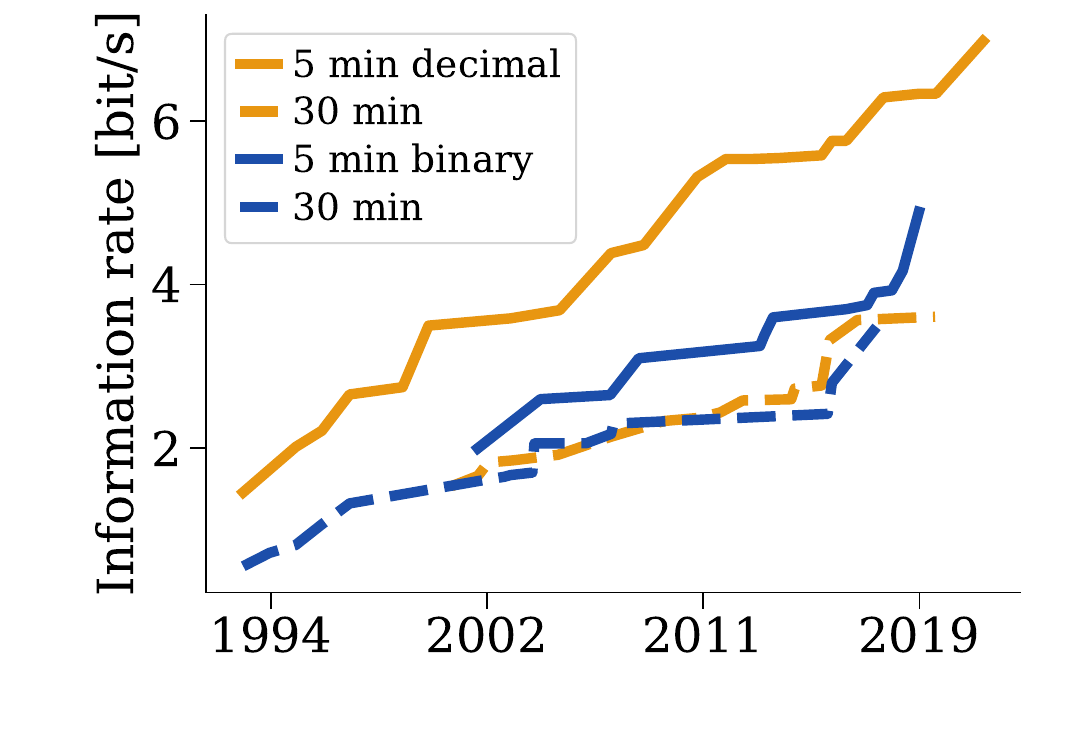}};
			\node[anchor=center, font=\normalsize\rmfamily] at (5.2,-5.3) {Decimal and binary digits};
			\node[anchor=north west] at (5,-5.5) {\includegraphics[width=0.335\textwidth]{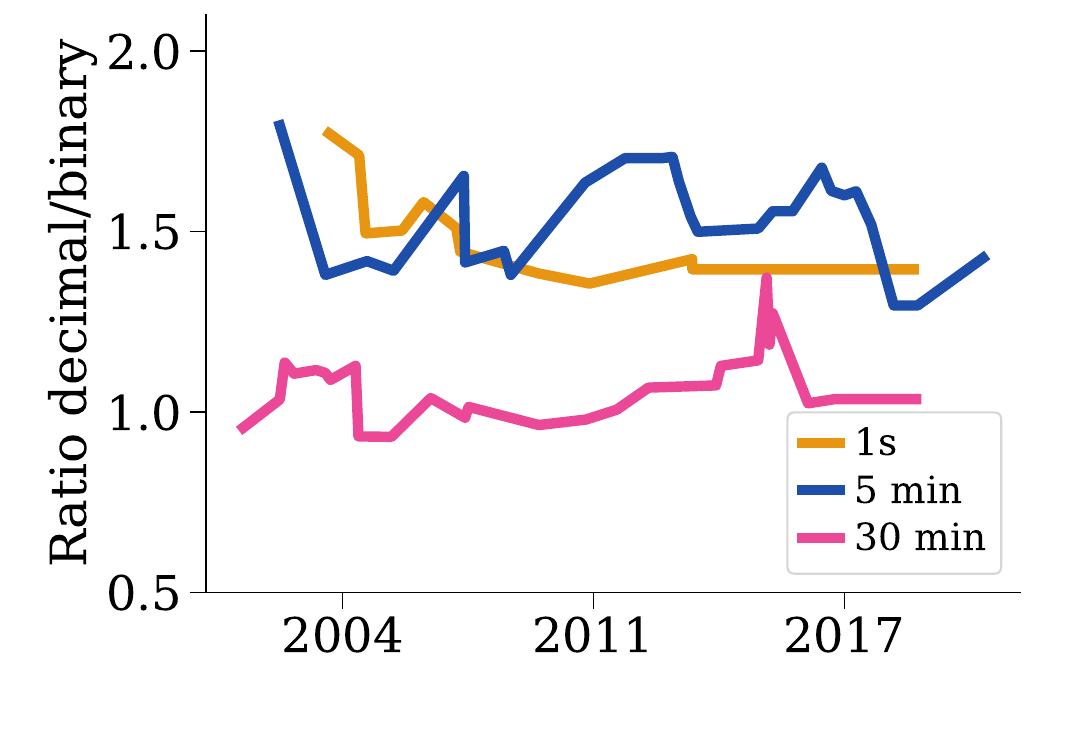}};
			\node[anchor=center, font=\normalsize\rmfamily] at (12.8,-5.3) {Decimal digits and cards};
			\node[anchor=north west] at (10,-5.5) {\includegraphics[width=0.335\textwidth]{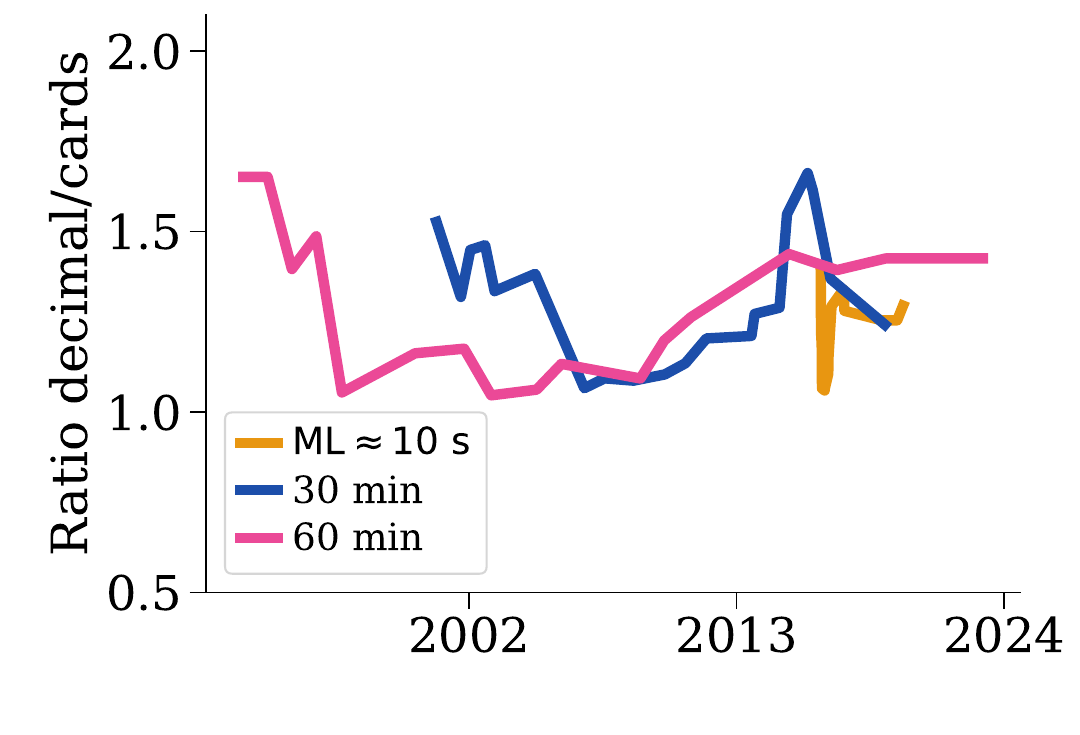}};
		\end{tikzpicture}
	}
	\caption{\textbf{A} Record development of selected disciplines since the inception of memory competitions. Left, all disciplines with five-minute memorization time. Middle, classical disciplines based on memorizing long sequences of decimal digits. Right, classical disciplines based on playing-card memorization.\\
		\textbf{B} Development of the information rate differences between decimal and binary digits, as well as cards. Left, record development of $5$ and $30$ min decimal- and binary-digits memorization. Middle, ratio of the information rates of all decimal and binary digits disciplines of comparable length. Right, ratio of the information rates of all decimal-digits and cards disciplines of comparable length. 
	} 
	\label{fig:record_developments_and_ratios}
\end{figure}

\unnumsubsection{Consistent differences between cards, decimal and binary digits}{subsec:optimal_encoding}
Classical competitions feature three symbolic coding systems: cards, decimal and binary digits. Over recent decades, world records have improved steadily with no clear indication that the trend is weakening (\Cref{fig:record_developments_and_ratios} \textbf{A}).
Despite this development, there are consistent relations between the achieved information rates across symbolic sets.

Decimal digits are processed faster than cards: this trend is convincing for disciplines from $10$ s to $60$ min memorization time over the past $30$ years (\Cref{fig:record_developments_and_ratios} \textbf{B}). The observed gap between cards and decimal digits suggests that different encoding schemes are processed at different speeds by humans. The $52$ options in playing cards add a complexity which is not present when memorizing decimal digits.

The comparison between decimal and binary digits is more complicated: there is a clear and consistent gap of around $40\%$ for the one-second and five-minute disciplines, but essentially the same memorization speed in the $30$ min discipline (\Cref{fig:record_developments_and_ratios} \textbf{B}).
The $40\%$ gap indicates that decimal digits strike a good balance between readability and complexity. We suppose that three factors explain the difference. First, the effect of word-length as more symbols are necessary to transfer the same information \autocite{word_length_effect}. For instance, more saccades might be necessary to accumulate the information or more peripheral vision is used, which could explain lower information rates \autocite{peripheral_vision}. Second, the effect of reduced salience as visual binary information is more redundant than decimal information \autocite{eye_movements}. Since binary saccade targets are less prominent, competitors might need to correct initial eye movements more often. Both effects can be experienced when reading the numbers
\[
350286 = 1010101100001001110_2,
\]
although the newest competition software tries to overcome the limitations through horizontal separation bars and colored cursor highlights \autocite{IAMtraining}. Third, competitors use systems for binary digits, which either carry less information per mnemonic image or were trained less. The most straightforward way to memorize binary digits is to convert three-digit binaries, say $010$, into a single decimal digit, here $2$. There are eight three-digit binary numbers, so that the strategy leads to a decrease of memorized information by 
\[
\frac{\log_2(8)}{\log_2(10)} \approx 0.9 \; .
\]
Competitors start by learning a system to memorize decimal digits, as decimal information is more ubiquitous in real life and featured in other disciplines in speed-memory.com and classical competitions. Binary-specific systems, overcoming the above factor, might not be trained at all or not to the same level of proficiency.

The assimilation of processing speeds in the $30$-min disciplines might be linked to the grading systems: decimal digits are presented in rows of $40$ ($\log_2(10^{40}) \approx 133$ bits), whereas binary digits are provided in rows of $30$ ($\log_2(2^{30}) = 30$ bits). If the row is correct $40$ respectively $30$ points are awarded. If a single digit is different, points are halved and no points are awarded if more than two digits are incorrect. As practically all mnemonic systems for decimal and binary digits encode at least two digits with an mnemonic image (\Cref{subsec:digits}), an error results in the loss of all points for the row. Thus, the grading system punishes an error in decimal digits with a deduction of $133$ bits and in binary digits only with $30$ bits. In the half-an-hour discipline forgetting and errors might become more important than the ability to parse information. This might force competitors to revise or spend additional time for decimal compared to binary digits.


\unnumsubsection{Associations are at the core of human memory}{subsec:associations}
The performances of mnemonists are frequently credited to their use of memory palaces. This undeniably is a major part of competitions and in fact the heavy focus on sequential tasks aims towards the use of memory palaces. 
It is still worthwhile pointing towards the disciplines where top performances do not involve the use of memory palaces: names and faces, historic dates and classical IAM images. The speed of historic dates the fastest among the five-minute disciplines (\Cref{fig:comparison_all_disciplines}). The objective is to memorize the year between $1000$ and $2100$ of fictional events (\Cref{subsubsec:historic_dates}). After memorization, the events are scrambled and competitors recall the respective years of the events. The current world record is $148$ by Prateek Yadav. If the events were not scrambled, his performance would correspond to memorizing a $\approx 450$ digit number, despite the fictional events not being optimized for associations. The events are almost complete sentences and often feature repetitions such as various queens, kings and princesses. We could interpret the discipline as providing an external memory palace. Any mental structure with an order, for instance, the sequence of events in your favorite movie, could function as a memory palace, a technique known as peg list \autocite{peg_list, memory_palace_as_peg_method}. Navigational routes are only the most ubiquitous such structure in our lives. Transforming the information into an mnemonic image or a concrete experience triggering our emotion, allows us to capitalize on our associative powers. Memory palaces are the most straightforward structure employed for sequential tasks, whereas associations are used in any discipline in memory competitions (\Cref{sec:techniques_and_records}).


\section{Closing remark}
Memory competitions have professionalized over the past three decades and will continue to amaze us. Rather than geniuses, ordinary people achieve astonishing skills through training. The statement ``we use only $10$ percent of our brain'' has been debunked by neuroscience. Scans show activity in all regions. To compete in nature, our bodies are highly efficient in minimizing calorie consumption: if inactive for a few weeks after a fracture, muscles will be gone. Our brain is being used, otherwise it would be minimized. Nevertheless, knowing what is possible in memory competitions, one could come to the conclusion ``we use our memory at only $10$ percent efficiency''.

%


\paragraph{Acknowledgements}
This work would not have been possible without the support of the memory sport community. I am grateful to Johannes Mallow and Simon Orton for sharing the dataset used to comment on past performances of top competitors during ML events. Johannes Mallow helped with several questions. Simon Orton kindly provided information on the generation of words and names in ML. I am thankful to Katie Kermode for discussions and for providing me information on the generation of names and words in the classical competitions.  Alex Mullen, Andrea Muzii, Don Michael Vickers and Enrico Marraffa generously gave insight into their performances. 

I would like to thank Markus Meister, who drew my attention to the information rates of human performances. Jan Rapp pointed out improvements of information calculations.
Andreas Herz and Martin Stemmler contributed with stimulating discussions and helpful feedback on the manuscript. 


\printbibliography

\appendix

\section{Classical and Memory League disciplines}

{ \def\spacingtabular{15pt}
	\begin{table}[H]
		\centering
		\footnotesize
		\begin{tabular}{C{15mm} C{50mm} C{18mm} C{20mm}}
			& memorization objective & memorization time [min] & grading scheme \\[5pt]
			\hline & & & \\[-1ex]
			decimal digits & a sequence of decimal digits & 5, 15, 30, 60 & rows of $40$  \\[\spacingtabular]
			binary digits & a sequence of binary digits & 5, 30 & rows of $30$   \\[\spacingtabular]
			auditory digits & a sequence of decimal digits, read aloud at one digit/s & up to 8 & counts until first error  \\[\spacingtabular]
			cards & a sequence of playing cards in the form of successive scrambled decks of $52$ playing cards & 10, 30, 60 & rows (decks)  \\[\spacingtabular]
			speed cards & the sequence of one scrambled deck of $52$ playing cards as quickly as possible & at most 5 & essentially perfect  \\[\spacingtabular]
			words & a sequence of words & 5, 15 & rows of $20$ \\[\spacingtabular]
			images & (WMSC) a sequence of abstract gray-scale images (IAM) a sequence of photos with arbitrary motifs (\Cref{fig:images_art_styles}) & 5 & rows of $5$ \\[\spacingtabular]
			historic dates & associate years between $1000$ and $2100$ to fictional events & 5 & deduction (guessing not allowed) \\[\spacingtabular]
			names and faces & associate international names to faces (first + last name) & 5, 15 & no deduction (guessing allowed) \\[\spacingtabular]
		\end{tabular}
		\caption{Overview of the disciplines in classical competitions. The disciplines were chosen to be easily corrected in the pen-and paper era and to be culturally independent. This explains the heavy focus on simple symbolic systems such as numbers and cards, where competitors can be expected to familiarize themselves with the symbol set.
			The first seven disciplines involve the memorization of long sequences. Typically they are graded in rows of $k$ items: a single error in a row leads to the row being counted as $k/2$, whereas with two or more errors no points are being awarded. 
			Speed cards is the only discipline in which the time rather then the amount counts. In recall, the number of cards counts until the first error - however, only an error-free performance leads to a competitive score.
			Historic dates and names and faces are associative disciplines. In recall, the events or the faces are in a random permutation and competitors need to reconstruct the year respectively the name. 
			The disciplines differ in our ability to estimate the information rate processed by humans in bit/s. For the purely number or card based disciplines entropies are easy to calculate. For images and historic dates we use theoretic lower-bounds, whereas for word and names we employ estimates based on database samples (\Cref{subsec:information_content_disciplines}).} \label{table:classical_disciplines}
\end{table}}

{ \def\spacingtabular{10pt}
	\begin{table}[H]
		\centering
		\footnotesize
		\begin{tabular}{C{20mm} C{70mm}}
			& memorization objective \\[5pt]
			\hline & \\[-1ex]
			words & a sequence of $50$ words \\[\spacingtabular]
			images & a sequence of $30$ photos with arbitrary motives, see \Cref{fig:images_art_styles} \\[\spacingtabular]
			numbers & an $80$ digit decimal number  \\[\spacingtabular]
			cards & the sequence of a scrambled deck of $52$ playing cards \\[\spacingtabular]
			national names & associate $30$ faces to national names (only first name) \\[\spacingtabular]
			international names & associate $30$ faces to international names (only first name) \\[\spacingtabular]
		\end{tabular}
		\caption{Overview of the disciplines in Memory League (ML). Compared to classical competitions the memorization time is shorter: at most one minute compared to at least five minutes in classical competitions with the exception of speed cards. Instead of optimizing the amount of information in a fixed time, competitors try to memorize a fixed-size set of information as quickly as possible.} \label{table:ml_disciplines}
\end{table}}

\section{Techniques and records in the different disciplines} \label{sec:techniques_and_records}
The following sections contain setting, strategies and world record performances for each discipline. 
Especially in short disciplines, competitors memorize the last few items/mnemonic images in short-term memory instead through links, a technique known as ``grab''. In ML and classical competitions this concerns only a small proportion of the information and it does not significantly alter the outcome of our calculations. The highest fraction is likely reached in ML images, where it is common to grab the sequence of the last four images, or $4/30 \approx 13\%$ of the information. In speed-memory.com higher fractions may be reached. We do not mention this again in the following sections.


\begin{figure}[!htb]
	\centering
	\hspace{1cm}\includegraphics[width=0.6\linewidth]{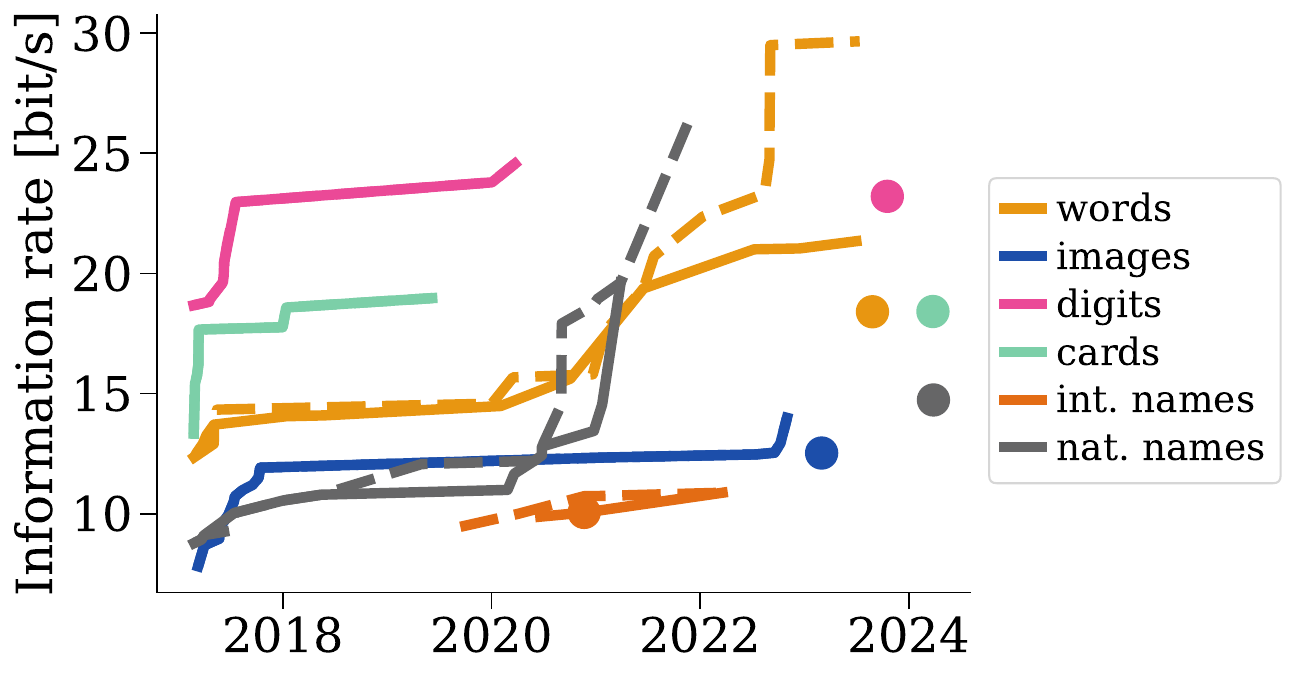}
	\caption{Memory League records. Unofficial and maximum entropy record development are shown as continuous and dashed lines, whereas the dots are the official records. We searched for scores with $100\%$ accuracy and maximum bit/s at $80 \%$ accuracy among performances of top competitors in a period from $2017$ to $2025$ (\Cref{subsec:data_acquisition}). For digits, cards and images, the $100\%$-accuracy records also maximized information processing. When competitors train, this remains invisible to other players, except when they choose to play against another competitor. Private training occurs at even higher speeds. Don Michael Vickers achieved $45/50$ words in $12.77$ s, equivalent to $42.04$ bit/s \autocite{donwordsyoutube}. By the time our data starts in $2017$ competitors had already practiced similar disciplines for two years, explaining the limited increase in a number of disciplines.} 
	\label{fig:comparison_memory_league_standard_and_with_errors}
\end{figure}

\unnumsubsection{Names and faces}{subsec:namesandfaces}
\paragraph{Setting and strategy} Faces with randomly assigned names are given during the memorization phase. Classical competitions feature first and last names, ML only first names. After memorization, the faces appear in a random order and the competitors need to recall the corresponding names. Classical names are always international, whereas in ML national and international names exist. The memorization time is $5$ min in national competitions and $15$ min in international and world championships. ML features $30$ faces and the objective is to remain error-free while ending the memorization period faster than the opponent.

People have tried to use memory palaces, but due to the reshuffling the top competitors instead focus on creating vivid stories for the names and connections to the faces (Strategy \ref{strategy_encoding} + \ref{strategy_associate}). Examples are names such as
\begin{align*}
	\text{Mary} & \hspace{1cm} \text{about to mary} \Rightarrow \text{imagine the person wearing a wedding dress},\\
	\text{James} & \hspace{1cm} \text{secret agent} \Rightarrow \text{imagine person as an agent}.
\end{align*}
Unfamiliar names are memorized by matching the contained letter combinations to more familiar sounds, words or structures such as a country-specific composition or ending.

		
		{\def\spacingtabular{10pt}
			\begin{figure}[!htb]
				\hfill
				\begin{minipage}{\textwidth}
					\centering
					\footnotesize
					\begin{tikzpicture}
						\node[anchor=north west] at (0, 0) { 
							\begin{tabular}{C{30mm} C{42mm} C{15mm} C{7mm} C{25mm}}
								& Description & WR & $[\frac{\text{bit}}{s}]$ & Record holder \\[2pt]
								\hline & & & & \\[-1ex]
								Classical names & Memorization time $15$ min & $224$ names & $4.88$ & Katie Kermode \\[\spacingtabular]
								& Memorization time $5$ min & $105$ names & $6.48$ & Katie Kermode \\[\spacingtabular]
								ML international names (official) & $30$ names & $58.51$ s & $10.05$ & Katie Kermode \\[\spacingtabular]
								ML international names (unofficial) & $30$ names & $54.1$ s & $10.87$ & Katie Kermode \\[\spacingtabular]
								& search results with $> 80\%$ accuracy for max bit/s & $54.1$ s & $10.87$ & Katie Kermode \\[\spacingtabular]
								ML national names (official) & $30$ names & $27.91$ s & $14.73$ & Matteo Cillo  \\[\spacingtabular]
								ML national names (unofficial) & $30$ names & $20.96$ s & $19.62$ & Matteo Cillo  \\[\spacingtabular]
								& search results with $> 80\%$ accuracy for max bit/s & $14.8$ s $29/30$ & $26.67$ & Jules Ballion\\[\spacingtabular]
							\end{tabular}
						};
					\end{tikzpicture}
				\end{minipage}
				\caption{World records in different names disciplines. Notably, the search for bit/s optimizing results was more successful for the national than the international names discipline in ML. Even for the top competitors in international names it is an achievement to memorize all $30$ names before the one minute is over, so there is no incentive to minimize time and thus maximize bit/s.}
				\label{fig:names}
		\end{figure}}
	
	\paragraph{Records} As an analogy, suppose a class contains thirty students. Then the top competitors in classical competitions take five minutes to memorize all names from three international classes. In ML the top competitors take less than a minute to memorize the names of one international class and even less than half a minute for a national class (\Cref{fig:names} \textbf{C}).
	
	The records among the different names categories demonstrates several effects: 
	the longer the memorization time, the slower the average speed in bit/s. Additionally, the difference between classical and ML scores, and official, unofficial and max bit/s scores can be explained by competitors having more often the chance to compete as well as being less risk-adverse in training and unofficial settings.
	ML national and ML international names scores reveal the effect of database size/names information: for the small national database ($< 2000$) competitors have ready-made associations for the overwhelming majority of names, whereas for international names their creativity is much more involved to come up with links for never-seen-before names \autocite{domain_expertise_chess, domain_expertise_wiktor}.

	Although the world record development has decelerated in recent years, it might be to early to draw conclusions. In classical competitions five disciplines involve digit memorization and seven disciplines involve sequence memorization. To maximize overall results, it might be more promising for competitors to improve their digits systems and memory palaces rather than familiarizing themselves even more with the structure of international names. 
		
		\unnumsubsection{Words}{subsec:words}
		\paragraph{Setting and strategies}
		The objective is to memorize a long sequence of words. In classical competitions, the sequence is subdivided into rows of $20$ words. A single error per row leads to the row being counted as ten points, and no points are awarded if there are two or more errors. In ML the sequence contains $50$ words and the best competitors optimize the time for perfect results.
		
		Any sequence of words can be memorized by forming a story. If the story becomes too long, it is faster and less risky to instead form shorter stories and place these at positions of a mental walk; the memory palace. This is often a well-known, real-life environment, such as a walk around your home, where one can easily make up a sequence of memorable positions: bed, desk, bathroom, ... .
		As the best competitors aim to memorize more than $100$ words in sequence in five minutes, or $50$ words in one minute in ML, memory palaces are the standard strategy in competitions. However, differences exist with respect to the number of words placed at a single locus. The most common is two, for instance, Katie Kermode and Don Michael Vickers are using this strategy (personal communication, \autocite{don}).

		
		{\def\spacingtabular{10pt}
			\begin{figure}[!htb]
				\hfill
				\begin{minipage}{\textwidth}
					\centering
					\footnotesize
					\begin{tikzpicture}
						\node[anchor=north west] at (0, 0) { 
							\begin{tabular}{C{30mm} C{42mm} C{15mm} C{7mm} C{25mm}}
								& Description & WR & $[\frac{\text{bit}}{s}]$ & Record holder \\[2pt]
								\hline & & & & \\[-1ex]
								Classical words & Memorization time $15$ min & $318$ words & $3.15$ & Katie Kermode \\[\spacingtabular]
								& Memorization time $5$ min & $145$ words & $4.3$ & Yanjaa Wintersoul \\[\spacingtabular]
								ML words (official) & $50$ words & $32.41$ s & $18.4$ & Don Michael Vickers \\[\spacingtabular]
								ML words (unofficial) & $50$ words & $27.95$ s & $21.34$ & Don Michael Vickers \\[\spacingtabular]
								& search results with $> 80\%$ accuracy for max bit/s & $18.5$ s $46/50$ & $29.66$ & Don Michael Vickers \\[\spacingtabular]
								& Youtube training \autocite{donwordsyoutube} & $12.77$ s $45/50$ & $42.04$ & Don Michael Vickers \\[\spacingtabular]
							\end{tabular}
						};
					\end{tikzpicture}
				\end{minipage}
				\caption{World records in the different words disciplines and additionally a training trial by Don Michael Vickers. }
				\label{fig:words}
		\end{figure}}
	
			\paragraph{Records} In classical competitions, competitors memorize less than half a word per second, whereas ML competitors have reached perfect memorization of two words per second for a much shorter period of time. The relationship between the achieved information rates shows similar effects as for names: shorter memorization time, more attempts, less risk-aware situations and pre-memorizable databases all contribute to higher information processing.
		
		\unnumsubsection{Images}{subsec:images}
		\paragraph{Setting and strategies}
		The objective of the images discipline is to memorize a sequence of randomly chosen motives. The imageset of WMSC, IAM and ML shows considerable differences.
		\begin{itemize}
			\item
			The discipline debuted in classical competitions with a memorization time of $15$ min in $2007$, although with images showing abstract, gray-scale, shapes of certain textures (\Cref{fig:images_art_styles} \textbf{A}). As competitors memorize hundreds of images, it is not feasible to search through a stack and assemble them in order. Images are instead presented in rows of fives and only the intra-row-sequence needs to be reconstructed. In recall, every row is in a random order and competitors indicate the previous positions by numbers $1-5$. A correct sequence of five images counts as five points, whereas an incorrect sequence leads to a deduction of a single point. WMSC competitions still feature this discipline nowadays.
			\item
			In IAM competitions, the abstract images discipline was replaced by the five-minute images discipline. The images depict clearly identifiable objects in different art styles (\Cref{fig:images_art_styles} \textbf{B}). The grading and recall system of rows of five was kept.
			\item
			Real-life images debuted in ML, where the objective is to memorize $30$ images in the correct sequence (\Cref{fig:images_art_styles} \textbf{C}). The interface allows to assemble the $30$ images in the original order during the recall time. 
		\end{itemize}
		
		{\def\spacingtabular{10pt}
			\begin{figure}[!htb]
				\centering
				\begin{minipage}{\textwidth}
					\centering
					\begin{tikzpicture}
						\node[anchor=north west] at (0, 0) { \includegraphics[width=0.6\textwidth]{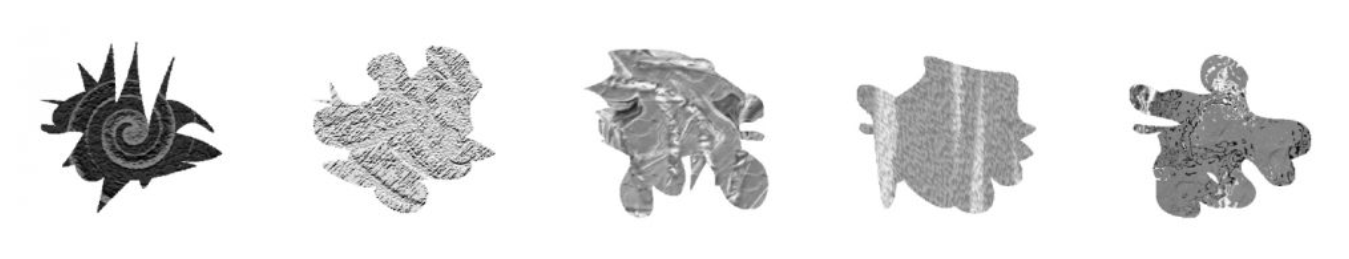}};
						\node[anchor=north west, font=\bfseries\large] at (-0.5, 0) {A};
					\end{tikzpicture}
				\end{minipage}
				\hfill
				\begin{minipage}{\textwidth}
					\centering
					\begin{tikzpicture}
						\node[anchor=north west] at (0, 0) { \includegraphics[width=0.6\textwidth]{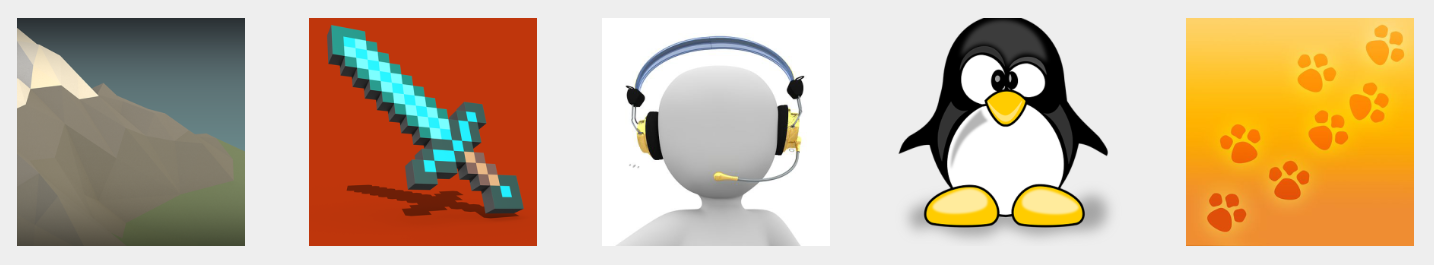}};
						\node[anchor=north west, font=\bfseries\large] at (-0.5, 0) {B};
					\end{tikzpicture}
				\end{minipage}
				\hfill
				\begin{minipage}{\textwidth}
					\centering
					\begin{tikzpicture}
						\node[anchor=north west] at (0, 0) { \includegraphics[width=0.6\textwidth]{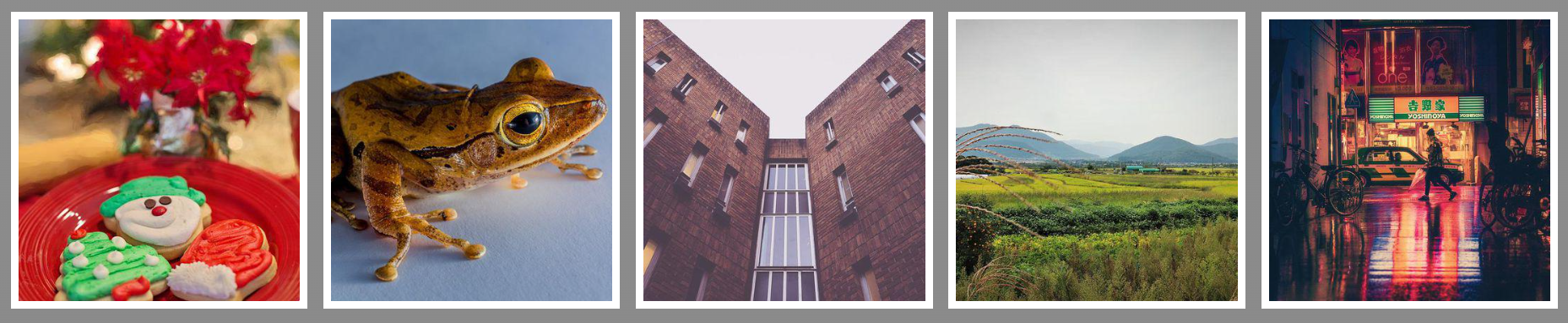}};
						\node[anchor=north west, font=\bfseries\large] at (-0.5, 0) {C};
					\end{tikzpicture}
				\end{minipage}
				\caption{Art style of images at the different competitions. \textbf{A} WMSC \textbf{B} IAM \textbf{C} ML }
				\label{fig:images_art_styles}
		\end{figure}}
	
		For abstract images competitors have invented encoding schemes to turn a complete row into a single mnemonic image, which in turn can be memorized by a memory palace \autocite{WMSC_abstract_images}. 
		For IAM images, competitors used loci initially. The current world record holder Enrico Marraffa just links the five images in one row to each other, saving the effort of a memory palace. The images are distinct enough to recognize the row during recall.
		In ML, competitors typically employ memory palaces, as the sequence is longer and the time scale shorter. Competitors prefer different numbers of items per loci, typically two or three. If only one image was placed per locus, the rate at which locations have to change would be substantial to memorize the order of $30$ images in $10$ s.
		
	\paragraph{Records} The quotient between results in $5$ min IAM images and $15$ min WMSC image is $ 2.03 \approx \frac{3.27}{1.61}$. The corresponding ratio is $\approx 1.34$ for names, $\approx 1.36$ for words and $\approx 1.45$ for digits. Unsurprisingly, this suggests that abstract images are harder to memorize than the IAM art-style. This illustrates that not only the information content is relevant for human competitors: storing a permutation is less information than recognizing the motives, but the latter is easier, if the motives are easily identified.
	
	{\def\spacingtabular{10pt}
		\begin{figure}[!htb]
			\hfill
			\begin{minipage}{\textwidth}
				\centering
				\footnotesize
				\begin{tikzpicture}
					\node[anchor=north west] at (0, 0) { 
						\begin{tabular}{C{30mm} C{42mm} C{15mm} C{7mm} C{25mm}}
							& Description & WR & $[\frac{\text{bit}}{s}]$ & Record holder \\[2pt]
							\hline & & & & \\[-1ex]
							Classical images (WMSC) & Memorization time $15$ min & $1048$ & $1.61$ & Huang Jinyao \\[\spacingtabular]
							Classical images (IAM) & Memorization time $5$ min & $711$ & $6.66$ & Enrico Marraffa \\[\spacingtabular]
							ML images (official) & $30$ images & $8.6$ s & $12.52$ & Matteo Cillo \\[\spacingtabular]
							ML images (unofficial) & $30$ images & $7.7$ s & $13.99$ & Matteo Cillo \\[\spacingtabular]
							& search results with $> 80\%$ accuracy for max bit/s & $7.7$ s & $13.99$ & Matteo Cillo \\[\spacingtabular]
						\end{tabular}
					};
				\end{tikzpicture}
			\end{minipage}
			\caption{World records in the different images disciplines. We have used different formulae to calculate the processed information of WMSC images, IAM images and ML images (\Cref{subsec:information_content_disciplines}).}
			\label{fig:images}
	\end{figure}}

		\unnumsubsection{Digits}{subsec:digits}
		
		We first explain the strategies and analyze the record development of decimal digits, before elaborating on the derived disciplines binary digits, auditive digits, historic dates and $\Pi$.
		
		\unnumsubsubsection{Decimal digits}{}
		
		\paragraph{Setting} Competitors are given a long sequence of decimal digits, in rows of $40$. If there is a single error in the row, the row counts $20$ points, otherwise no points are awarded. In ML, the objective is to memorize an $80$-digit number in under one minute.
		
		\paragraph{Strategies}
		
		The main idea is to map the digits onto prememorized mnemonic images, for example objects or people, and instead memorize their sequence in a memory palace.
		
		\textit{Major system} The most straightforward system is to memorize a map for single digits, say $0$ = ring, $1$ = candle, $2$ = swan, ... . Of course, such a system quickly becomes repetitive. The problem can be solved by creating one-hundred, or one-thousand mnemonic images to memorize two or even three decimal digits at once. A new problem arises: how can one initially commit such a large system to memory? To facilitate learning, one can assign letters to the digits $0-9$. The most prominent example is the Major System in \Cref{table:major_system}, which associates a certain consonant or group of similarly sounding consonants to every single digit \autocite{major_system}. Vowels do no not encode digits, thus giving flexibility in creating words, which match a certain two- or three-digit number. For instance, the combination $094$ could be memorized as Zebra,
		\[
		094 = S/Z B R = \text{Zebra}
		\]
		by utilizing the code below.
		
		{ \def\spacingtabular{8pt}
			\begin{table}[!htb]
				\centering
				\begin{NiceTabular}{c c c c c c c c c c}
					0 & 1 & 2 & 3 & 4 & 5 & 6 & 7 & 8 & 9 \\[2pt]\hline 
					& & & & & & & & &  \\[-1ex]
					s, z & t & n & m & r & l & ch & c,k & f,w & p,b 
				\end{NiceTabular}
				\caption{The initial memorization of the mapping from digits to mnemonic images is facilitated by systems such as the major system. } 
				\label{table:major_system}
		\end{table}}

		\textit{Person Action Object (PAO)} The second major approach follows a more modular strategy. A combination of a memorable person, a corresponding action and item is associated with every two-digit number. To name a few examples, $02$ could be snowman, to melt and carrot; $36$ could be a famous basketball player, to slam dunk and a basketball; $82$ a sports fan, to scream and a flag. A complete list of one hundred such combinations suffices to memorize six-digit decimal numbers. The first two digits encode the person, the middle two digits the action and the last two the object:
		\[
		\underbrace{02}_{\text{Person}} \underbrace{36}_{\text{Action}} \underbrace{82}_{\text{Object}} = \text{Snowman slam dunks flag}.
		\]
		Note that snowman encoding $02$ capitalizes on the above major system. Of course, also a PA or PO system is possible.
		
		{ \def\spacingtabular{8pt}
			\begin{table}[!htb]
				\centering
				\footnotesize
				\begin{NiceTabular}{C{10mm} C{20mm} C{20mm} C{18mm} C{18mm}}
					& $\#$mnemonic images & Bits / mnemonic image & Learnability & Brute-force \\[2pt]
					\hline & & & & \\[-1ex]
					2-digit & $100$ & $6.64$ & &  \\[\spacingtabular]
					PAO & $3 \times 100$ & $6.64$ & &  \\[\spacingtabular]
					3-digit & $1000$ & $9.97$ & &  \\[\spacingtabular]
					4-digit & $10 \times 1000$ & $13.29$ & &  \\[\spacingtabular]
					\CodeAfter
					{\tikz\draw[thick, <-] (3-|4.5) -- (6-|4.5);}
					{\tikz\draw[thick, <-] (3-|5.5) -- (6-|5.5);}
				\end{NiceTabular}
				\caption{A brief overview of the most common systems used to memorize digits and additionally the $4$-digit system, most prominently used by Simon Reinhard \autocite{simondigits}.} \label{table:digits_systems}
		\end{table}}
		
		\textit{Comparison of the different systems}
		All systems, even a simple two-digit one, can be trained to great speed surpassing any non-systematic approach to numbers memorization. PAO and the three-digit system are most commonly used by top competitors. The systems lay emphasis on different aspects. The larger the system, the more information is conveyed by a single mnemonic image and the more variable stories become. However, more time needs to be spend to create the images and to achieve the same encoding speed. In case of an error, competitors ``brute-force'' by going through all possible combination until there is a subjective ``click'' of having found the correct mnemonic image. This strategy becomes harder as the number of mnemonic images increases. When placing several images on one location, there is a risk of swapping images. The three PAO mnemonic images on a single locus carry additional sequential information, reducing this risk.
		
\begin{figure}[!htb]
			\centering
			\includegraphics[width=0.5\linewidth]{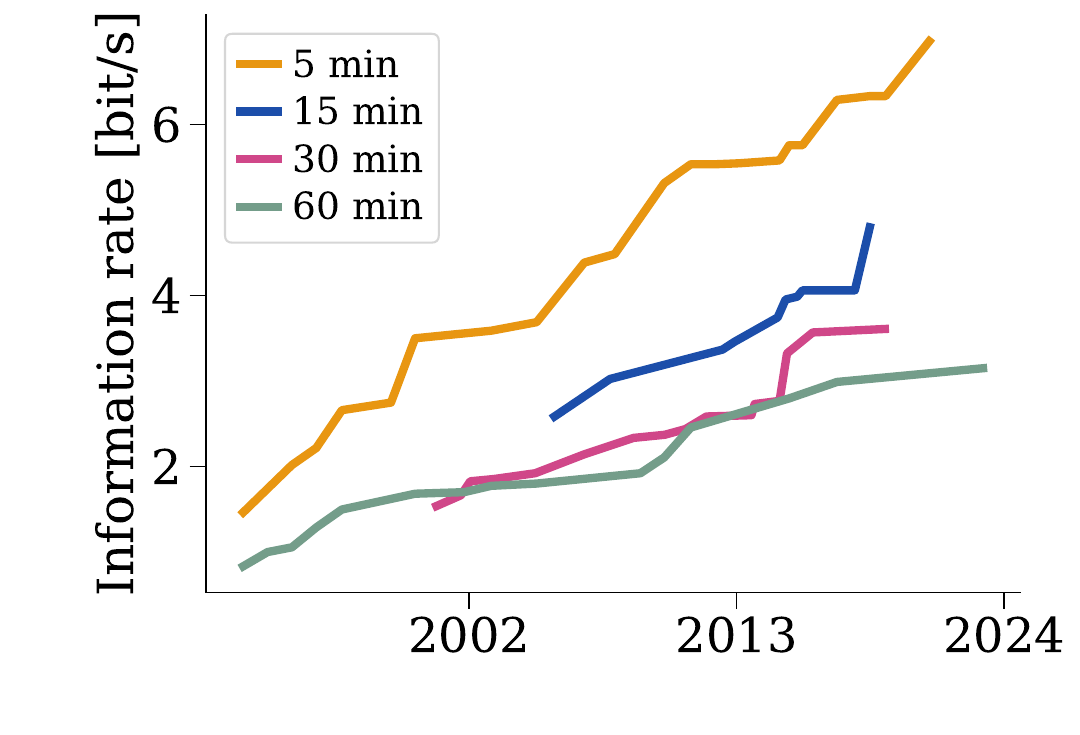}
			\caption{
					The decimal digits world records have improved consistently since the inception of memory competitions.
				}
			\label{fig:comparison_digits}
		\end{figure}

		\paragraph{Records}
		The performances in classical decimal digits have consistently improved over time (\Cref{fig:comparison_digits}). The quotient of the performances in disciplines of different time has also remained approximately constant, suggesting a time-scale to human performance law (\Cref{fig:power_law_and_model}).

		
		{\def\spacingtabular{10pt}
			\begin{figure}[!htb]
				\hfill
				\begin{minipage}{\textwidth}
					\centering
					\footnotesize
					\begin{tikzpicture}
						\node[anchor=north west] at (0, 0) { 
							\begin{tabular}{C{25mm} C{42mm} C{15mm} C{9mm} C{25mm}}
								& Description & WR & $[\frac{\text{bit}}{s}]$ & Record holder \\[2pt]
								\hline & & & & \\[-1ex]
								Classical digits & Memorization time $60$ min & $3412$ & $3.15$ & Orkhan Ibadov \\[\spacingtabular]
								& Memorization time $30$ min & $1955$ & $3.61$ & Sylvain Arvidieu \\[\spacingtabular]
								& Memorization time $15$ min & $1300$ & $4.8$ & Munkshur Narmandakh \\[\spacingtabular]
								& Memorization time $5$ min & $630$ & $6.98$ & Andrea Muzii \\[\spacingtabular]
								ML digits (official) & $80$ decimal digits & $11.45$ s & $23.2$ & Alex Mullen \\[\spacingtabular]
								ML digits (unofficial) & $80$ decimal digits & $10.82$ s & $24.56$ & Andrea Muzii \\[\spacingtabular]
								& search results with $> 80\%$ accuracy for max bit/s & $10.82$ s & $24.56$ & Andrea Muzii \\[\spacingtabular]
								& Youtube training \autocite{andreadigitsyoutube} & $9.75$ s 80/80 & $27.26$ & Andrea Muzii \\[\spacingtabular]
								Speed Memory (official) & Memorization time $4$ s & $31$ & $25.75$ & Joaquin Garcia \\[\spacingtabular]
								& Memorization time $1$ s & 21 & $69.76$ & Ram\'on Campayo \\[\spacingtabular]
								Speed Memory (unofficial) & Memorization time $0.5$ s & $17$ & $112.95$ & Ram\'on Campayo \\[\spacingtabular]
							\end{tabular}
						};
					\end{tikzpicture}
				\end{minipage}
				\caption{World records in the different digits disciplines and additionally a training trial by Andrea Muzii. }
		\end{figure}}

		\unnumsubsubsection{Binary digits}{subsubsec:binary}
		\paragraph{Setting and strategies} Binary digits are only featured in classical competitions and presented in the same manner as decimal digits, except the use of rows of $30$. Having a system for decimal digits, the most straightforward way to memorize binary digits is to convert three-digit binaries, say $010$, into a single decimal digit, here $2$, and memorize the decimal sequence instead. 
		Few competitors have taken up the effort to create more elaborate encoding schemes. Notably, Ben Pridmore created a system which covers all ten-digit binary numbers with a single mnemonic image, which creates a similar information density $\log_2 (2^{10}) = 10 \approx 9.97 = \log_2(1000)$ as the standard three-digit number system \autocite{bensystem}.
		
		\paragraph{Records} The comparison between the world records in decimal and binary digits shows that the information rate for binary digits is consistently lower, at least in the one-second and five-minutes disciplines. We explain this phenomenon in \Cref{subsec:optimal_encoding}.
		
		{\def\spacingtabular{10pt}
		\begin{figure}[!htb]
		\hfill
			\begin{minipage}{\textwidth}
				\centering
				\footnotesize
			\begin{tikzpicture}
				\node[anchor=north west] at (0, 0) { 
				\begin{tabular}{C{25mm} C{42mm} C{15mm} C{9mm} C{25mm}}
	& Description & WR & $[\frac{\text{bit}}{s}]$ & Record holder \\[2pt]
	\hline & & & & \\[-1ex]
	Classical binary & Memorization time $30$ min & $6270$ & $3.48$ & Munkshur Narmandakh \\[\spacingtabular]
	& Memorization time $5$ min & $1467$ & $4.89$ & Munkshur Narmandakh \\[\spacingtabular]
	Speed Memory & Memorization time $4$ s & $100$ & $25$ & Joaqu\'in Garcia \\[\spacingtabular]
	& Memorization time $1$ s & $50$ & $50$ & Joaqu\'in Garcia \\[\spacingtabular]
\end{tabular}
				};
			\end{tikzpicture}
			\end{minipage}
			\caption{World records in the different binary digits disciplines.
			}
			\label{fig:comparison_binary_decimal}
		\end{figure}}
		
		\unnumsubsubsection{Auditive numbers}{}
		\paragraph{Setting}
		Competitors listen to a track in which a human voice announces one decimal digit/s. After the end of the track, competitors recall the sequence. To determine the score, the number of correct digits is counted only until the first error. There are three attempts during competitions (for instance, $100$, $200$ and $500$).
		The unique characteristic of this discipline is that the memorization speed is the same for all competitors and, whereas all other disciplines are based on vision, this discipline is auditive. Training regimes center on practicing at speeds faster than one decimal digit/s.
		
		\paragraph{Records}
		The current world record $456$ of Lance Tschirhart corresponds to seven and a half minutes one digit/s without error \autocite{IAMrecords}. Notably, the current average speed of the one-hour digits marathon is $0.95$ digit/s, a similar speed. This illustrates that it is difficult for human beings to perform error-free memorization (\Cref{subsec:no_machines}).
		
		
		\unnumsubsubsection{Historic dates}{subsubsec:historic_dates}
		\paragraph{Setting and strategy}
		Competitors memorize fictional historic dates, together with a year between $1000$ and $2100$, for example,
		\[
		1453 \text{ Princess captured by elephants.} 
		\]
		After the memorization period, the fictional events are in random order with the objective of recalling the year. The unique feature of the discipline is that numbers are involved, but no memory palace is required, as one can directly link mnemonic image and the corresponding information.
		Fictional dates are necessary to ensure that competitors do not profit from prior historic knowledge. As there are only $1100$ possible years, the first digit $1$ or $2$ is less important for this task. This allows to memorize the middle digit in a redundant fashion. As $1100$ combination can be covered by a complete system of mnemonic images (for instance, three-digit + additional two-digit system), this seems to be the most efficient strategy.
		\paragraph{Records} The bit/s rate achieved in this discipline is the highest among classical disciplines with five minutes memorization time. If one were to give the historic events in the same order in recall, the current world record of $148$ by Prateek Yadav corresponds to memorizing a roughly $450$ decimal digit number without using a memory palace (\Cref{subsec:associations}).
		
		\unnumsubsubsection{\texorpdfstring{$\Pi$}{Π}}{subsubsec:pi_setting_performance}
		\paragraph{Setting and strategies} Compared to the classical format and ML, in $\Pi$ competitions the time during memorization is unlimited and recall is often in oral form.
		The unlimited time allows one to use an encoding scheme such as \Cref{table:major_system} and create a particularly fitting story for the digits of $\pi$ by, for example, varying the length of words. Interestingly, this renders the use of memory palaces less important. With enough memorization time, long vivid stories are possible and the need to break the information rapidly into chunks is less great \autocite{piphilology}.
		
		\paragraph{Records}
		The official world record by Suresh Kamar Sharma is at $70030$ digits, equivalent to around $29$ kilobyte \autocite{pi}. We compare this record to the standard disciplines in \Cref{subsec:power_law}.
		
		\unnumsubsection{Cards}{subsec:cards}
		\paragraph{Setting}
		Competitors memorize one deck of $52$ playing cards in the speed-cards discipline and ML cards. Real playing cards can only be chosen in traditional competitions. In the longer classical disciplines, competitors are confronted with several decks, which serve the role of rows in the other disciplines. A full deck counts as $52$ points, a single error leads to $26$ and otherwise no points are awarded. The memorization time is either $10, 30$ or $60$ min. In speed cards there is a five-minute period during which competitors can start their trial. Each competitor is timed separately. Timing is provided by the software or, for real cards, by a timer, which is started and stopped with both hands. Recall starts collectively after the five-minutes are over. Note, however, that competitors will start revising and recalling immediately after stopping their timer even without the actual recall starting.
		\paragraph{Strategies}
		Similar to decimal digits, the core of systems for the cards discipline are encoding schemes. Due to the structure of $52$ playing cards, they can be even more elaborate then for decimal digits.
		 
		\textit{One card}
		A map between the $52$ playing cards and $52$ mnemonic images suffices to memorize cards. It is possible to reach times below $30$ s just based on this system. Nowadays, the top competitors have moved to either PAO or two card systems.
		
		\textit{Person Action Object (PAO)}
		The system for numbers generalizes in a straightforward manner to cards, as there are $100$ two-digit numbers, but only $52$ playing cards. By creating a translation of cards into two-digit numbers, one can transfer the system and memorize blocks of three cards by a single modular mnemonic image. 
		
		\textit{Two card} To memorize two cards from a single deck at once, one requires a total of $2652$ distinct mnemonic images. Competitors such as Ben Pridmore have mastered this system \autocite{bensystem}. Johannes Mallow proposed the following innovation to cut the number of required mnemonic images in half: two combinations of pairs of cards, one with a red card first and one with a black card first, are mapped to the same mnemonic image. Competitors distinguish which of the combinations comes first by encoding the binary information in the number of items per loci. If a red-first combination appears, competitors jump to the next loci, whereas if a black-first combination comes next, competitors append the mnemonic image to the story at the current locus.
		To facilitate the creation of either system, competitors typically reuse the $1000$ mnemonic images of their three-digit system \autocite{lance_two_card}.
		
				{ \def\spacingtabular{8pt}
			\begin{table}[!htb]
				\centering
				\footnotesize
				\begin{NiceTabular}{C{10mm} C{20mm} C{20mm} C{20mm} C{18mm} C{18mm}}
					& $\#$mnemonic images & Mnemonic images per deck & Average bits/image & Learnability & Brute-force\\[2pt]
					\hline & & & & & \\[-1ex]
					1-card & $52$ & $52$ & $4.34$ & &  \\[\spacingtabular]
					PAO & $3 \cdot 52$ & $17 \cdot 3 + 1 = 52$ & $4.34$ & &  \\[\spacingtabular]
					Zo\'n System & $169 + 256 = 425$ & $13 \cdot 3 = 39$ & $5.78$ & &  \\[\spacingtabular]
					Half-2-card & $\frac{52\cdot 51}{2} = 1326$ & $26$ & $7.68$ & &  \\[\spacingtabular]
					2-card & $52 \cdot 51 = 2652$ & $26$ & $8.68$ & &  \\[\spacingtabular]
					\CodeAfter
					{\tikz\draw[thick, <-] (3-|5.5) -- (7-|5.5);}
					{\tikz\draw[thick, <-] (3-|6.5) -- (7-|6.5);}
				\end{NiceTabular}
				\caption{A brief overview of the most common systems used to memorize cards and additionally an interesting, intermediate system invented by Jan Zo\'n \autocite{janzoncards}. The Zo\'n system encodes four cards with three mnemonic images, one for the $4^4 = 256$ combination of the four cards colors and two for the $13^2 = 169$ combinations for the values of two cards each. The half-2-card system manages to store $26$ bit in the interaction with the route, rather than the mnemonic images. A full deck of cards corresponds to $\log_2(52!) = 225.58$ bit.} \label{table:cards_systems}
		\end{table}}
	
		\textit{Comparison of the different systems}
		Similar to numbers, the systems lay different emphasis on the aspects of information density, learnability and the ability to the recover from partial memory loss. As the variable length of stories on loci is manageable, the half two-card system sacrifices only one bit per mnemonic image for much easier learnability.
		
		{\def\spacingtabular{10pt}
			\begin{figure}[!htb]
				\begin{minipage}{\textwidth}
					\centering
					\footnotesize
					\begin{tikzpicture}
						\node[anchor=north west] at (0, 0) { 
							\begin{tabular}{C{30mm} C{42mm} C{15mm} C{7mm} C{25mm}}
								& Description & WR & $[\frac{\text{bit}}{s}]$ & Record holder \\[2pt]
								\hline & & & & \\[-1ex]
								Classical cards & Memorization time $60$ min & $1829$ & $2.21$ & Orkhan Ibadov \\[\spacingtabular]
								& Memorization time $30$ min & $1202$ & $2.9$ & Sylvain Arvidieu \\[\spacingtabular]
								& Memorization time $10$ min & $550$ & $4.02$ & Andrea Muzii \\[\spacingtabular]
								& One deck, $52$ playing cards & $12.74$ s & $17.71$ & Shijir-Erdene Bat-Enkh \\[\spacingtabular]
								ML cards (official) & One deck, $52$ playing cards & $12.25$ s & $18.41$ & Alex Mullen \\[\spacingtabular]
								ML cards (unofficial) & One deck, $52$ playing cards & $11.89$ s & $18.97$ & Alex Mullen \\[\spacingtabular]
								& search results with $> 80\%$ accuracy for max bit/s & $11.89$ s & $18.97$ & Alex Mullen \\[\spacingtabular]
							\end{tabular}
						};
					\end{tikzpicture}
				\end{minipage}
				\caption{World records in the different cards disciplines. }
				\label{fig:cards}
		\end{figure}}

	\paragraph{Records} As speed cards are the only discipline comparable to the corresponding ML discipline, it is reassuring to find similar performances. The comparison to decimal digits in \Cref{fig:record_developments_and_ratios} \textbf{B} shows that the information rate achieved in cards is consistently lower. This indicates that the symbolic complexity of playing cards renders them harder for human competitors, even though a potential error in cards can be more easily recovered by brute-forcing the combinations of the remaining cards of the deck (\Cref{subsec:optimal_encoding}).

	It remains to be seen which system prevails in the long run. The most recent official world record in speed cards of $12.74$ s by Shijir-Erdene Bat-Enkh has been set with PAO, whereas Alex Mullen set a time of $12.25$ in ML with the half two-card system.

\section{Data analysis} \label{sec:data_analysis}
\unnumsubsection{Data acquisition for world record performances}{subsec:data_acquisition}
To analyze classical competitions, we used the world record histories \autocite{IAMrecords} from the statistics website \autocite{IAMstatistics} of the International Association of Memory (IAM), where also the official online world records for ML can be found \autocite{ml_records}. Results are only counted as world records if the result is obtained during an online competition with videos recorded from two different angles. Separate in-person records are kept but are outdated as in-person competitions have ceased to exist for ML since the COVID-19 pandemic. The world records and development for the $1$ s and $4$ s decimal and binary disciplines can be found on speed-memory.com \autocite{speed_memory_records, speed_memory_record_development}.

Johannes Mallow (commentator on ML competitions \autocite{johannestwitch}, former world champion) and Simon Orton (ML developer) provided us with a database of $191503$ results of top competitors from the recent results on the ML website \autocite{MemoryLeague} covering the period between June $2017$ and January $2025$. The dataset is used by commentators to assess the likelihood of participants to get a certain score during online competitions. The unofficial word record performances were generated by searching for the best results with $100\%$ accuracy at any time. To search for the results with maximum information, we considered results with at least $80\%$ accuracy (for instance, $> 0.8 \cdot 80 = 64$ digits correct).


\unnumsubsection{Information content of the different disciplines}{subsec:information_content_disciplines}
Our calculations are based on Shannon's entropy, defined as
\[
H (\mathbb{P}) = - \sum_{x \in \Omega} \log_2 ( \mathbb{P}(x)) \mathbb{P}(x)
\]
for a probability distribution $\mathbb{P}$ on a discrete underlying event space $\Omega$. 
Marking rules and the high number of possibilities prevent guessing of competitors. Therefore, if we calculate the entropy associated to the recalled items, it will correspond to transmitted information. Ideally, we would prefer to have a distribution of answers of a competitor. The records present only one realization, but at least the disciplines consist out of a sequence of repetitive microtasks such as individual digits. For all disciplines, our calculations are based on the idea to provide a lower bound on the information necessary to explain the human performance. 

For example, in classical digits, cards, words and images disciplines we assume performances are perfect, even though the row grading scheme leads to raw scores which are lower than the number of correctly recalled items. The disciplines are given in rows of $k$ items. A single error in a row leads to the row being counted as $k/2$, whereas with two or more errors no points are being awarded. If the last row is only recalled up to a certain item, indicating that the competitor has not memorized further, the rule is applied to the partial row.

For words and names, we use estimates based on large samples of databases. For the associative aspect in names, images and historic dates, we use a lower bound, which avoids any assumptions on human cognition. We do not incorporate any of the underlying mental processes, such as the memory palace, in our calculations. We denote entropy associated to achieving a raw score $n$ in a discipline by $H ( \text{discipline}, n)$.

\paragraph{Names and faces}
Faces with a first and a last name are provided during the memorization phase in classical competitions. In the recall period, faces are scrambled and competitors need to recall the names. A point is awarded for every correctly retrieved name, so two points are possible per face. This implies that a raw score of $n$ corresponds to memorizing at least $n/2$ faces. ML national and international names feature $30$ faces with only a first name.

IAM names are generated as follows:
\begin{itemize}
	\item
	Names are uniformly chosen from eight different `regions', which cover different language branches.
	\item 
	In five-minute names, there are $12$ ``long'' names of eight letters or more and in the $15$ min version $24$ long names. The long names are chosen uniformly among first and last names.
\end{itemize}
In the memorization phase the number of names equals the world record plus $20\%$. As the current five-minute world record is $105$, we assumed a proportion $12/126$ of long names. Katie Kermode, who co-developed the latest IAM training software \autocite{IAMtraining}, provided us with a list of the numbers of short/long, male/female per region. This allowed us to calculate the entropy carried by IAM names as
\begin{align*}
	H (\text{First name}) = 13.34 \;,\\
	H(\text{Second name}) = 12.24 \;.
\end{align*}
A score of $n$ requires at least $n/2$ faces to be distinguished in recall. Assuming that each of the faces was instead a number, or as clearly distinguishable as possible,  one could use the following strategy to solve the discipline. During memorization, one would first assemble the names in a canonical order or sort by corresponding number and then memorize the names as a list. In recall, one could recreate the order and then recall the list of names. Having a number for each face, one can see that the lower bound for the information of this strategy and thus association tasks is $\log_2 (\# \text{faces})$. Almost all of the provided information is memorized in record performances, allowing us to assume a similar amount of first and last names. We obtain the overall lower estimate
\[
H (\text{Classical names}, n) \approx n \Big(\frac{H(\text{First name}) + H(\text{Second name})}{2} + \log_2 (n /2) \Big).
\]
ML does not share the precise names distribution, but Simon Orton, the developer of ML, kindly applied the entropy formula for us. Both male and female names have an entropy of
\[
H (\text{ML national name} ) = 8.8 \;.
\]
This allows us to calculate the entropy associated to a national names performance in ML as
\[
H (\text{ML national names}, n) \geq n \big( H(\text{ML national name}) + \log_2 (n) \big)
\]
for $0 \leq n \leq 30$. International names consists out of all $41$ national names databases combined. Simon Orton provided us again with the result of the entropy formula
\[
H(\text{ML international name}) = 14.7 \;.
\]
Similar to national names, we computed the formula
\[
H(\text{ML international names}, n) \geq n \big( H(\text{ML international name}) + \log_2(n) \big)
\]
for $0 \leq n \leq 30$.

\paragraph{Words}{}
Words is the only discipline for which no database for IAM competitions exists. As competitors memorize words in their native language, it is easier for organizers to provide translations for the then limited task set. As such, performances are slightly harder to compare than in other disciplines. Nevertheless, the discipline can be trained on the IAM training website \autocite{IAMtraining}. We assume that players would have complained if the training mode was too easy and used a list provided by Katie Kermode. It contains $1683$ concrete nouns, $829$ abstract nouns and $763$ verbs with some words classified for several categories. The composition of the sequence of words in competitions is $80\%$ concrete nouns, $10\%$ abstract nouns and $10\%$ verbs \autocite{Rulebook}. Accounting for words being in several categories, this leads to the estimate
\begin{align*}
	H(\text{Classical word}) = 8.9 \;.
\end{align*}
We then use the formula
\[
H (\text{Classical words}, n) \approx n H(\text{Classical word}).
\]

Simon Orton informed us that ML words are uniformly chosen among the ML words database. We found more than $3900$ different words, giving us the lower bound
\[
H(\text{ML word}) \geq \log_2(3900) \approx 11.93 \; .
\]
This allows us to calculate the entropy associated to a words performances in ML as
\[
H (\text{ML words}, n) \geq n H(\text{ML word}) = n \cdot 11.93 \; .
\]
for $0 \leq n \leq 50$, as $50$ words are given.

\paragraph{Images}
In this discipline, the task is to recreate the original order of a sequence of images.
While the scores reach several hundred  in the classical format, it is important to note by what system the images discipline is marked. Images are presented in rows of five, and in recall each row of five images is shown in a random order and participants need to indicate the previous order by numbers. There is a deduction of one point per incorrect row preventing guessing. 
In the abstract images disciplines at WMC competitions, people have managed to just store the permutation applied to the five images per row (\Cref{subsec:images}) implying the formula
\[
H(\text{Classical images (WMC)}, n) \geq \frac{n}{5} \log_2 (5!).
\]
Enrico Maraffa, the world record holder in IAM images, confirmed in personal communication that he is not using memory palaces, but only links. Therefore, he would be able to recall the same information if the rows were scrambled. As he is able to identify the rows, similar to the associative aspect in names, we add $\log_2 \big(\frac{n}{5} \big)$ for the row identification to arrive at
\[
H(\text{Classical images (IAM)}, n) \geq \frac{n}{5} \Big( \log_2 (5!) + \log_2\big( \frac{n}{5} \big)  \Big).
\]

In ML, participants see $30$ images and need to reassemble them during recall. Thus, the entropy of the ML images discipline is 
\[
H(\text{ML images}, n) \geq \log_2(n!) 
\]
for $0 \leq n \leq 30$.

\paragraph{Digits}{}
The calculation of the number of bits conveyed by decimal digits and binary digits is straightforward:
\begin{align*}
	H (\text{Binary digits} , n) &= n,\\
	H(\text{Decimal digits} , n) &= n \log_2(10),\\
	H(\text{ML digits},n) &= n \log_2 (10), \hspace{1cm} \forall 0 \leq n \leq 80.
\end{align*}

\paragraph{Historic dates} The objective is to memorize the year between $1000$ and $2100$ of fictional historic dates. In recall, the fictional events are in a random order without years. 
The entropy carried by the year of a single memorized historic date is $\log_2 (1100) = 10.1$. We again need to account for the associative component by adding $\log_2 (n)$. In reality encoding will be less efficient. We can thus provide a lower bound to the information processed in the discipline by
\[
H(\text{Historic dates} , n ) \geq n \big( \log_2 (1100) + \log_2 (n) \big).
\]

\paragraph{Cards}{}
The entropy of a subset of $52$ playing cards is
\[H(\text{One deck, } n \text{ out of 52 cards}) = H (\text{ML Cards}, n) = \log_2 \Big( \frac{52!}{(52- n)!} \Big).\]
We again assumed that memorization was perfect and that any remainder $\mod 52$ originates from the memorization of a partial deck at the end of the recalled sequence, leading to the formula
\[
H (\text{Classical cards}, n) = \Big\lfloor \frac{n}{52} \Big\rfloor \log_2 (52!) + \log_2 \Big( \frac{52!}{(52 - n \mod 52)!} \Big).
\]

\section{Probabilistic model} \label{sec:probabilistic_model}
We present a short probabilistic model relating the information-rate power law to the probability of perfect memorization in short tasks. The main idea is that competitors try to remain error-free and that a long discipline can be decomposed into shorter, independent tasks. It necessarily neglects many of the aspects described in the previous sections. We model one of the sequential disciplines. Suppose that competitors try to achieve perfect memorization to avoid incurring any of the rows being awarded no points. To do so, they aim for perfect memorization with a certain probability
\[
\mathbb{P} (\text{all correct}) \approx p \in (0,1).
\]
We subdivide the sequence into groups of $k$ items, each with entropy $H(\text{item})$. Let $A(k,t)$ be the event that a competitor attempts and correctly recalls $k$ elements in $t$ seconds memorization time excluding the reading time $r$ for the group. Let $T$ be the overall time of the discipline in seconds. Assuming that groups of $k$ elements are independent of each other, we could then write
\begin{align*}
	p \approx \mathbb{P} (\text{all correct}) = \mathbb{P} \big(A (k,t) \big)^{T/(t + r)} \Rightarrow \mathbb{P} \big( A (k,t) \big) \approx p^{(t +r) /T}.
\end{align*}
The official records suggest that the information rate scales as a power law of the discipline, so we obtain
\[
a T^b = R(T) = \frac{k H(\text{item}) \cdot \frac{T}{(t + r)}}{T} = \frac{k H (\text{item})}{(t + r)} \Rightarrow T = \Big(\frac{k H (\text{item})}{a(t + r)} \Big)^{1/b}.
\]
This implies that
\begin{equation*}
	\mathbb{P} \big( A (k,t) \big) = p^{ (t +r) \Big(\frac{a(t + r)}{k H (\text{item})} \Big)^{1/b} }. 
\end{equation*}
The prediction $\mathbb{P} \big( A (k,t) \big)$ for fixed $k$ can be found in \Cref{fig:power_law_and_model} \textbf{B}. If, alternatively, one fixes $t$, the model predicts that the probability to memorize all $k$ items resembles a reversed logistic growth curve. For a small number $k$ of items the probability is close to one. As the number of items $k$ increases, the decay of the probability is first slow, accelerates and then decelerates, so that the probability asymptotically converges to zero.

\end{document}